\documentclass[aps,pre,reprint,groupedaddress, showpacs]{revtex4-1}

\usepackage{amssymb}
\usepackage{amsfonts}
\usepackage{amsmath}
\usepackage{bm}
\usepackage{graphicx}
\newcommand{\V }[1]{\bm{#1}}

\newcommand{\be}{\begin{equation}}
\newcommand{\ee}{\end{equation}}
\newcommand{\bee}{\begin{eqnarray}}
\newcommand{\eee}{\end{eqnarray}}

\begin{document}

\title{The impact of clogging on the transfer dynamics of hydraulic networks}


\author{D. S. Griffani} 
\email[]{danielle.griffani@sydney.edu.au}
\author{P. Rognon} \author{I. Einav}
\affiliation{Particles and Grains laboratory, School of Civil Engineering, The University of Sydney. Building J05, 2006 Sydney Australia}

\date{\today}

\begin{abstract}
We investigate how clogging affects the transfer properties of a generic class of materials featuring a hydraulic network embedded in a matrix. 
We consider the flow of a liquid through fully saturated hydraulic networks which transfer heat (or mass) by advection and diffusion, and a matrix in which only diffusion operates.
Networks are subjected to different clogging scenarios (or attacks), changing their microstructure and flow field.
A series of canonical cooling tests are simulated using a tracer method prior to and following attack. Results quantify the threat posed by different types and degrees of attack to a system's transfer properties. An analytical framework is introduced to predict this vulnerability to attacks, rationalised in terms of their effect on the physical mechanisms underlying the transfer dynamics.
\end{abstract}


\maketitle

\section{Introduction}
Hydraulic networks often serve as an effective pathway to distribute heat, nutrients or oxygen into a low permeability matrix. 
For instance, fractured networks in rocks \cite{berkowitz2002characterizing, geiger2010non}, vascular networks in plants and animals \cite{jensen2013physical, corson2010fluctuations, katifori2010damage, sack2006leaf}, and channel networks in microfluidic heat exchangers and engineered tissue serve this role \cite{pence2010simplicity, miller2012rapid, hu2014efficient}. 
The microstructure of these networks present very different geometrical and topological characteristics. Furtherstill, natural and anthropogenic attacks can clog or render certain channels incapable of carrying a flow, leading to a redistribution of the flow through the network coupled with a rapidly changing microstructure \cite{guariguata2012jamming, campbell2010jamming, freund2013flow, valdes2006particle, valdes2007particle, ivanov2008applications, baveye1998environmental, rinck2000interrelationships, depreitere2006mechanics, shipley2010multiscale, bebber2007biological, katifori2010damage, choat2012global, loepfe2007relevance, singurindy2005role, andre2006influence, bachler2005coupled}.
How these effects translate to impact the rate of transfer between the matrix and the liquid flowing through the network is largely unknown.

Models predicting the flow in hydraulic networks are relatively well established. Considering a fully saturated network and laminar flow conditions, the macroscopic flow rate $q$ [m/s] through the network satisfies a Darcy law $q = - \frac{k}{\eta}\nabla P$ involving the pressure gradient $\nabla P$ [Pa/m], the fluid viscosity $\eta$ [Pa.s] and the permeability $k$ [m$^2$].  
The permeability, is expressed by the Kozeny-Carman model as $k\propto R^2\phi/T^2 $ \cite{fuerstman2003solving, costa2006permeability, jivkov2013novel}. This model reflects the role of only the porosity $\phi$ and channel radius $R$ [m] aspects of the network geometry; meanwhile, the role of the network topology, or branching pattern, is embedded solely in the tortuosity $T$. The tortuosity represents a ratio of two distances: the average distance travelled by the fluid in the network from point $A$ to point $B$, and the Euclidian distance $AB$ \cite{duda2011hydraulic, koponen1996tortuous, rognon2014explaining}. 

The dynamic of transfer of heat or mass between the matrix and the network is controlled by two processes: the advection within the network and the diffusion within the matrix islands bordered by the channels.  Double Porosity, Multiple INteracting Continua and Matrix Diffusion models are examples of continuum formulations that successfully capture these processes via the following 
macroscopic quantities: the Darcy velocity $q$ for the advection; and the matrix diffusivity $D_0$ [m$^2$/s] and (typical) matrix island size $\ell$ for the diffusion \cite{ xu2001modeling, pruess1982practical, haggerty1995multiple, carrera1998matrix}. According to these models, for a given Darcy flow and matrix diffusivity, the transfer dynamic is entirely determined by the size $\ell$, while other network geometrical and topological features do not need to be known.

The question is whether the geometrical and topological characteristics embedded in the macroscopic quantities $q$ and $\ell$ are sufficient to capture the effect of different attacks on the transfers, and if so, how do these quantities evolve as a network is subjected to increasing degrees and different types of attack?

In a recent study \cite{rognon2016vulnerability}, we established how attacking networks by clogging a number of selected channels may strongly affect the flow of liquid. Firstly, clogging always affects the Darcy flow $q$ by lowering the number of free flowing channels and thus lowering the effective porosity $\phi$. Secondly, clogging may modify the network tortuosity by redirecting the flow path. Attacks resulting in the clogging of channels transverse to the Darcy flow lead to a slight decrease in tortuosity, favouring more direct flow paths. By contrast, attacks that constrict the flow path in the macroscopic flow direction were found to dramatically increase the network tortuosity. While the dynamics of transfer between the matrix and the network are expected to be affected by these types of attacks and the change in flow paths they induce, there is no established framework to rationalise and predict this effect.  

In this Letter, we seek to establish how channel clogging affects the ability of hydraulic networks to transfer heat or mass from/to the matrix they are embedded in; and how vulnerable this ability is to different types and extent of attack. To this aim, we use a tracer method to simulate a series of canonical cooling tests highlighting the dynamics of transfer for networks with differing geometries and topologies, before and after being subjected to different clogging scenarios. 
 
\section{Simulating flow and transfer in matrix embedded hydraulic networks} \label{sec:method}
\begin{figure}
 \includegraphics[width=\columnwidth]{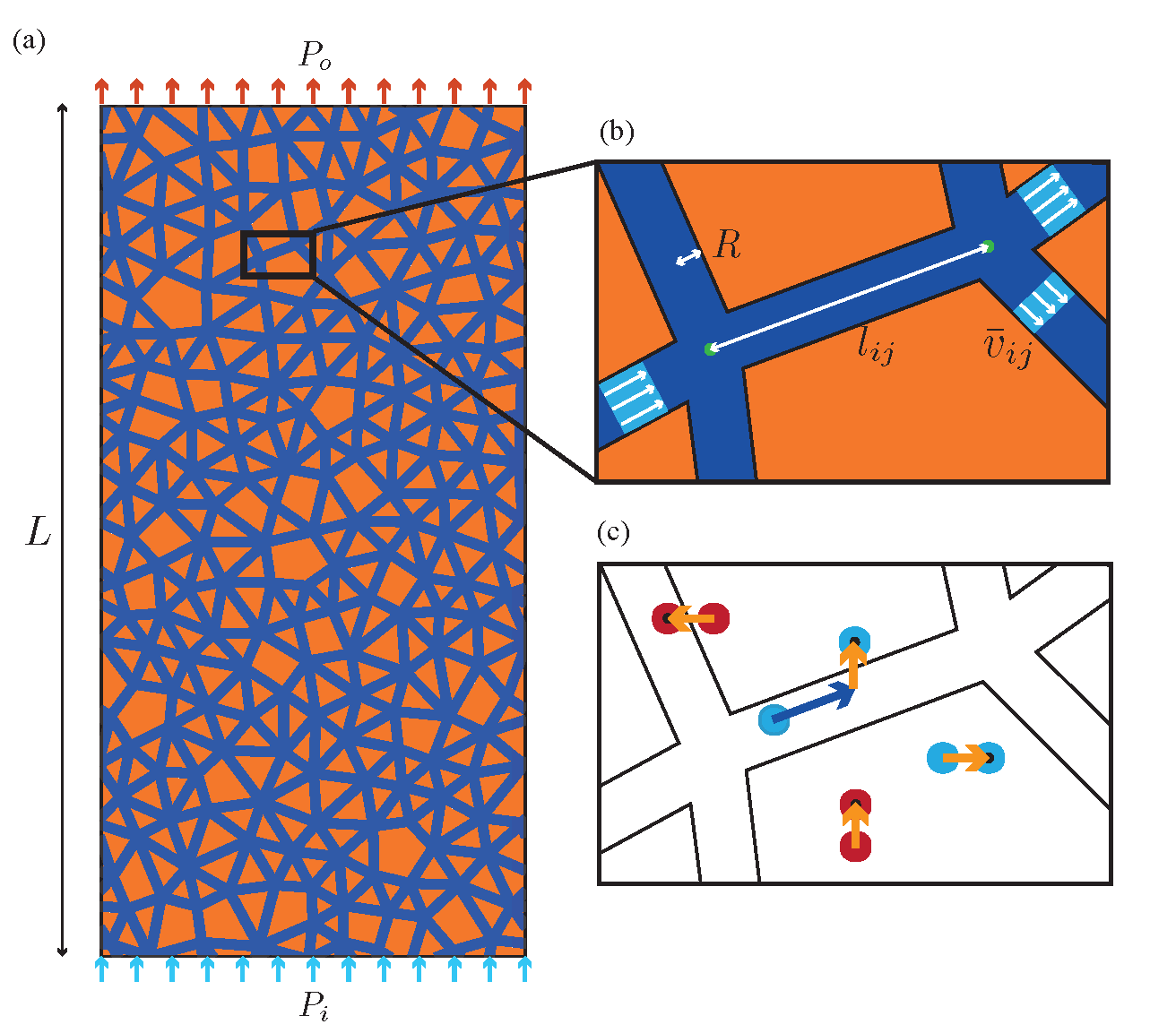}
\caption{Simulated example: (a) hydraulic network comprised of approximately 500 channels embedded in a diffusive matrix. The system is bi-dimensional and periodic in both directions. (b) The local flow field (c) Typical tracer motions at a given time step, including random walk (orange arrow) and advection (blue arrow). Tracers carry a scalar indicator of temperatures  $T_{0}$ (red) and $T_{in}$ (blue). Temperature in a domain is the average scalar of all the tracers it contains. 
\label{fig1:system} }     
\end{figure}

We consider systems comprised of a matrix pervaded by a network of connected channels of similar radius $R$ (see Fig. \ref{fig1:system}). The network is fully saturated by an incompressible fluid. The fluid only flows within the network (and not the matrix), driven by a macroscopic pressure gradient ($\nabla P$) in the y-direction. Transfer occurs by advection in the network, and by diffusion in both the network and the matrix.  For simplicity, the diffusivity $D_0$ in the network and in the matrix are equal. 

Hydraulic networks are constructed by randomly placing nodes within the simulation domain whilst enforcing a minimum distance of $5R$ between any two nodes, and connecting selected pairs of nodes by a channel. Following  \cite{rognon2014explaining}, network topologies including (i) \textit{Lattice} networks; (ii) \textit{Gabriel} networks; (iii) \textit{Epsilon} networks; and (iv) \textit{K-nearest} networks, are obtained by applying different rules of connection.

The flow field in each channel of the networks is numerically calculated using the method detailed in  \cite{rognon2014explaining,rognon2016vulnerability}. The method considers a laminar Hagen-Poiseuille flow in each channel, and the corresponding linear relationship between the average flow velocity $\bar v_{ij}$ within the channel and the pressure gradient along the channel:
 \be \label{eq:av_velo}
 \bar v_{ij} = - \frac{R^2}{8\eta} \nabla p_{ij},
 \ee
 
\noindent where $\nabla p_{ij} =\frac{p_j-p_i}{l_{ij}}$ is the pressure gradient along the channel length $l_{ij}$ and $p_i$ and $p_j$ are the pressures at the two extremities of the channel (or nodes) (see Fig. \ref{fig1:system}b). The liquid viscosity $\eta$ was set at $1\times 10^{-3} Pa.s$ for all simulations. 

The pressure at each node is deduced from the mass budget at each node, which defines a linear system of equations of the form $\V M \cdot {\V P} = \V {S} $; where $\V{P}$ is a vector containing the pressures at all nodes,  $\V {S}$ is a vector containing the known pressure drop for channels crossing the periodic boundaries in the y direction, and $\V {M}$ is a square matrix built from for the mass budget at each and every node. The pressures at all the individual nodes can be deduced from ${\V P} = \V{M}^{-1}\cdot \V {S} $, which requires the numerical inversion of $\V M$. Having found the pressures for every pair of end nodes defining a channel $p_i$ and $p_j$, the channel flow velocities are then deduced from Eq. (\ref{eq:av_velo}).

This method does not account for the viscous dissipation at the channel junctions (nodes), which can safely be neglected for slender channels $| l_{ij} |\gg R$, a condition that is enforced for all channel network constructions. 
In principle, this method could be applied to solve for the flow in a network of porous channels, where the flow within individual channels satisfy a Darcy law of the form  $\bar v_{ij} = - \frac{k_{ij}}{\eta} \nabla p_{ij}$, where $k_{ij}$ is the permeability of the channels \cite{walsh1981effect,latham2013modelling,carey2015fracture}. 
 
Diffusion in the matrix and in the network, and advection in the network are simulated using a tracer method as in  \cite{griffani2013rotational,griffani2014transfer}.
In principle, the method involves placing a large number of passive numerical tracers at random positions throughout the entire channel-matrix domain, and integrating their motion over short time steps $dt$ (see Fig. \ref{fig1:system}c). The motion of tracers includes a random walk component: at each time step each tracer is assigned a random velocity vector of constant norm $v_0=\sqrt{\frac{4D_0}{dt}}$, the orientation of which is selected at random from the four possible orthogonal directions of a square grid. As a result of the random walk, the tracers exhibit a diffusive behaviour described by the diffusivity $D_0$ \cite{schirmacher2015random}, and can move across the channel-matrix boundary by diffusion. 
To simulate advection, the local flow field is superimposed upon the random walk. Thus, tracers within a channel are also subjected to a laminar flow characterised by the average flow velocity for that channel $\bar v_{ij}$. Our previous work \cite{griffani2014transfer} showed that for the same average flow velocity, the nature of the flow profile across the channels does not significantly impact the effective transfer properties of a system. Accordingly, for simplicity, a plugged flow field was imposed in all channels as illustrated in figure \ref{fig1:system}b.

The scheme used to integrate the tracer motion comprises an Eulerian first-order integration of the flow of tracers inside a channel followed by the random motion. This scheme is applied to each tracer. The time step, $dt$ adopted for each simulation is a fraction of the lesser of the smallest typical diffusion time scale and advection time scale, that is $dt=\frac{1}{100}min\left(\frac{R^2}{D_0}, \frac{min(l_{ij})}{ max(\bar{v_{ij}})}\right)$ [s]. The tracer density, $\rho$ is set according to $\rho=\frac{2}{R^2}$ [tracers/$m^2$]. We systematically checked that smaller time steps and/or more tracers yield the same results.

\section{Probing the transfer dynamics}\label{sec:test}
The canonical cooling test performed on each system involves defining a homogeneously hot initial state, where the temperature of the matrix and in the channels are equal (temperature $ T_0$), and injecting a cooler liquid (temperature $T_{in}$) at the inlets of the channel network at a prescribed rate. 
All throughout the experiment, both the inlet temperature and injection rate are kept constant. The average temperatures of the matrix  $\bar T_s(t)$ and channels $\bar T_l(t)$ are monitored over time, from which we define the normalised temperature variations in the matrix and in the channels as follows:
\be
\Delta \bar T_s(t) = \frac{ \bar T_s(t) -T_{in}}{  T_0 -T_{in}}; \;\Delta \bar T_l(t) = \frac{ \bar T_l(t) -T_{in}}{  T_0 -T_{in}}.
\ee

As in \cite{griffani2014transfer}, we use the \textit{transfer efficiency} $E$ to quantify the cooling dynamics of a given system, which is defined as:
\be \label{eq:E}
E=\frac{t_{diff}}{t_{80}}; \; t_{diff}=\frac{L^2}{D_0}
\ee

\noindent where $t_{80}$ is the time needed to cool the matrix by $80\%$ and $t_{diff}$ a typical diffusion time across the system length, $L$. 

\section{Transfer dynamics in undamaged networks}

The cooling test was performed on systems with different network topologies, geometries and matrix diffusivities $D_0$, and subjected to different flow rates $q$.  The network geometry is characterised by the porosity $\phi$ and a typical inter-channel distance, or effective matrix island size $\ell_e$. $\ell_e$ scales with the average length of the channels, $\bar l$ according to $\ell_e=a\bar\ell$ where $a$ is a constant. 

Accordingly, the systems are governed by a set of three dimensionless numbers including porosity $\phi$, a P\'eclet number $Pe$ and dimensionless island size $\lambda$ defined as:
\be \label{eq:peclet}
\lambda=\frac{L}{\ell_e},\; Pe=\frac{qL}{D_0}.
\ee

\noindent  $\lambda$ compares the typical inter-channel distance to the system length. The P\'eclet number compares a typical diffusion time scale $t_{diff}=\frac{L^2}{D_0}$ to a typical advection time scale $t_{adv}=\frac{L}{q}$ across the system length, and thus measures the relative rate of these two modes of transfer. 

\begin{figure*}
\centering
\includegraphics[width=1.5\columnwidth]{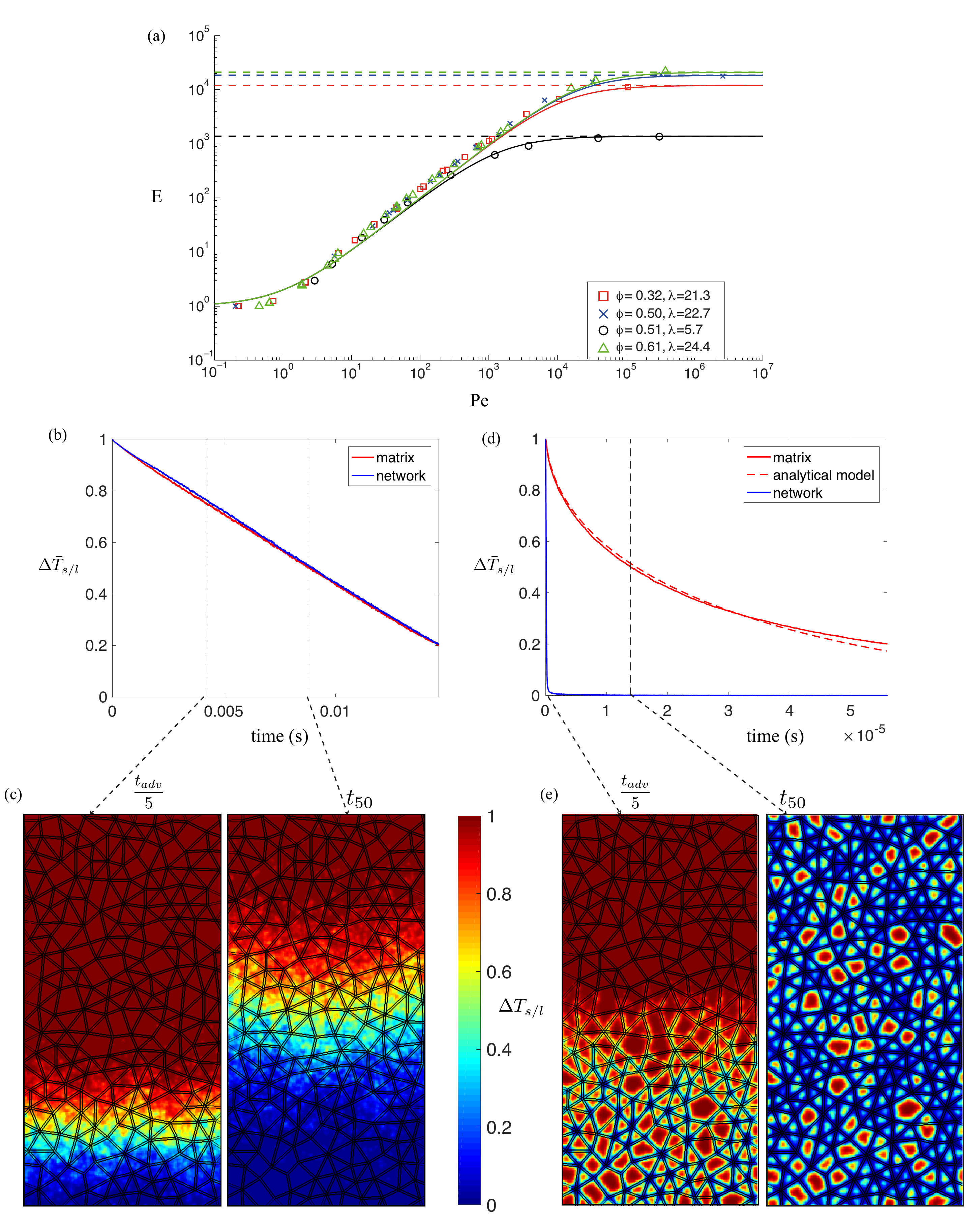}
\caption{Cooling dynamics of undamaged Gabriel networks. (a) Scaling of the cooling efficiency with the P\'eclet number: each symbol corresponds to cooling test simulated with a different triplet $Pe$, $\phi$ and $\lambda$. Dashed horizontal lines denotes the maximum efficiency obtained by using the $Pe_\infty$ method (see text). Solid lines represent the predictions of the model in Eq. \ref{Eq:model_E} (b,d) cooling dynamics of systems in regime II and III, respectively, simulated with the tracer method ($\phi=0.5$, $\lambda =22.8 $, (b) $Pe = 62$ and (d) $2.6\times10^6$.); in (d), the dashed line corresponds to prediction of the analytical model (\ref{eq:dirichlet_2d}), where $\ell_e=0.52\bar{l}$, truncating the sum to $n, m <= 19$. (c,e) snapshots of the temperature field during the cooling test.
\label{fig2:undam} }    
\end{figure*}

Figure \ref{fig2:undam}a shows the measured efficiencies corresponding to a Gabriel network for various values of P\'eclet number $Pe$, porosity $\phi$  and dimensionless island size $\lambda$.  These cooling dynamics exhibit three regimes. For low P\'eclet numbers ($Pe \lesssim 1$ - Regime I), the efficiency is close to one, indicating that the cooling dynamic is mainly controlled by the process of diffusion across the system. For intermediate P\'eclet numbers (Regime II), the efficiency increases linearly with with the P\'eclet number ($E\approx Pe$). Figures \ref{fig2:undam}b,c show that, in this regime, the (average) temperature of the matrix nearly equals the temperature of the liquid in the network, which both scale like $\Delta \bar T_{s}(t) = \Delta \bar T_{l}(t) \propto- \frac{t}{ L/q}$, indicating an advection driven cooling. Figure \ref{fig2:undam}c further indicates that cooling results from the progression of an approximately uniform and regular temperature front  

For larger P\'eclet numbers, the efficiencies seem to reach a constant, maximum value $E_\infty$ and the cooling exhibits a dynamic that is in stark contrast with regime II's. Figures \ref{fig2:undam}d,e show that, whilst the temperature of the liquid in the network is still advection driven and scales like $\Delta \bar T_{l}(t) \propto - \frac{t}{ L/q}$, the decay of the matrix temperature is much slower. This leads to a pattern of hot matrix islands surrounded by cold liquid.

We implemented a specific type of simulation referred to as $Pe_\infty$ to evidence the processes underpinning this dynamic. With the $Pe_\infty$ method, the liquid in the channels is set at a constant temperature $T_{in}$ at all times, which represents a situation where flow in the network is infinitely fast, instantly flushing out any heat coming from the matrix. The simulation then only considers diffusion in the matrix, with Dirichlet boundary conditions at the interface between the matrix and the channels. Figure \ref{fig2:undam}a and d show that the efficiency and transfer dynamics measured with this method corresponds to those measured at finite values of P\'eclet number in regime III. This indicates that the cooling dynamics in regime III are governed by a process of diffusion across the matrix islands referred to henceforth as, `micro-diffusion'. 

Considering islands that are square in shape, with a size $\ell_e$, and subjected to Dirichlet boundary conditions, their diffusion-driven cooling would satisfy the analytical solution of the diffusion equation:
\bee \label{eq:dirichlet_2d}
\Delta \bar T_s(t) &=& w^{-1}  \sum_{n \, odd}^{\infty}\sum_{m \, odd}^{\infty}   \frac{1}{n ^2}   \frac{1}{m ^2}\exp \left(-\frac{t}{t_{n,m}} \right)  \\
w &=& \sum_{n\,odd}^{\infty} \sum_{m \, odd}^{\infty}    \frac{1}{n ^2}\frac{1}{m ^2}\\
t_{n,m} &=& \frac{\ell_e^2}{D_0} \frac{ 1 }{\pi^2 (n^2+m^2)} 
\eee

\noindent Figure \ref{fig2:undam}d shows that by selecting an appropriate value for the effective island size of the undamaged network, $\ell_e=0.52\bar{\ell}$, this model captures the cooling dynamics in regime III, confirming that it is driven by the mechanism of micro-diffusion.

In the appendix, we show that the proposed analysis for the Gabriel network captures the cooling dynamics measured for all four network types. The effects of the topological variations on the transfer dynamics can be completely resolved by accounting for the change in Darcy velocity $q$ and island size $\ell_e$ induced by the variations. This implies that at least for relatively homogeneous topologies like those considered, the role of the network topology does not significantly affect the nature of the transfer dynamics which is governed by $Pe$ and $\lambda$.

\section{Transfer dynamics in clogged networks}
In \cite{rognon2016vulnerability}, we classified a number of clogging mechanisms including jamming of particulate fluids in channels \cite{guariguata2012jamming, campbell2010jamming, freund2013flow}, bioclogging \cite{ivanov2008applications, baveye1998environmental, rinck2000interrelationships}, vein rupture and channel damage \cite{depreitere2006mechanics, shipley2010multiscale, bebber2007biological}, cavitation \cite{choat2012global, loepfe2007relevance}, precipitation and calcification \cite{singurindy2005role, andre2006influence, bachler2005coupled} into two main attack scenarios. In the first scenario, referred to as \textit{Major} attack, channels with the highest pressure gradient and highest flow rate are preferentially clogged. In the second scenario, referred to as \textit{minor} attack, channels with the lowest pressure gradient and lowest flow rate are preferentially clogged. 
For both scenarios, we further distinguished \textit{block} attacks in which targeted channels are clogged all at once, from \textit{iterative} attacks in which channels are clogged one by one. With the iterative attacks, the flow in the network is updated each time a channel is clogged, leading to a new flow and pressure distribution in the network. The next channel to be clogged is selected from this updated state, resulting in a progressive clogging of channels.  

A system featuring an undamaged Gabriel network (statistically similar to that illustrated in Fig. \ref{fig2:undam}) was subjected to a series of each of the four types of attack: \textit{major iterative (MI), major block (MB), minor iterative (mi) and minor block (mb)}. The proportion of channels clogged in the network, $\mathcal{P}$ was varied for each of the types of attack up to a maximum $\mathcal{P}$ that would result in at least one node becoming disconnected from the network. 
 Following each attack, the change in porosity, tortuosity and permeability of the network followed similar relationships with $\mathcal{P}$ to those reported in \cite{rognon2016vulnerability}. The range of these variables resulting from the entire spectrum of attacks performed on the undamaged network are presented in table \ref{tab:parameter} 

\begin{table} [htp]
\begin{center}
\begin{tabular}{c c c  c}
\hline
$\mathcal{P} $ & $\phi $& $T$ &$\frac{k}{k_0}$\\
\hline
$0 \rightarrow 40\%$& $ 0.32 \rightarrow 0.50  $&$ 1.19\rightarrow 2.57 $&$ 0.26\rightarrow 1.08$ \\
\hline
\end{tabular}
\end{center}
\caption{Range of porosity, tortuosity and permeability normalised by the undamaged system permeability $k_0$ encountered following network attacks.  
\label{tab:parameter} 
}
\end{table}
Two example macroscopic pressure gradients $\nabla P$ were applied to the system following each attack. The pressure gradients selected were those utilised in figure \ref{fig2:undam}, the lowest and highest of which yielded transfer dynamics in regime II and regime III respectively for the undamaged system. $L$ and $D_0$ were also kept the same as the undamaged system in figure \ref{fig2:undam}.

\begin{figure}
\centering
\includegraphics[width=\columnwidth]{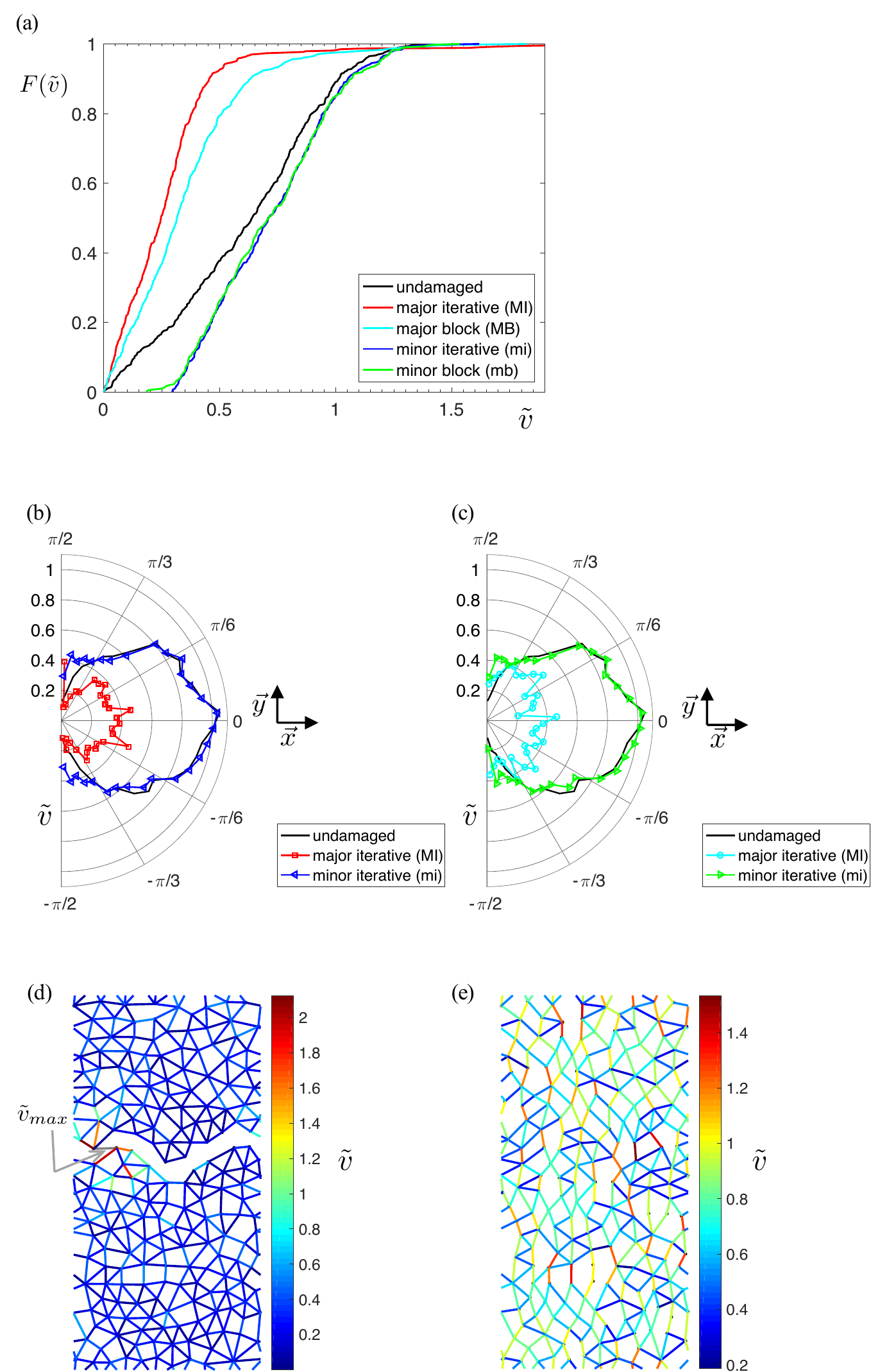}
\caption{Flow in the network before (undamaged) and after examples of each of the four types of attack: major iterative (MI), major block (MB), minor iterative (mi), and minor block (mb). The proportion of channels clogged, $\mathcal{P}$ in all but the MI attack is $20\%$. $\mathcal{P}= 2.8\%$ for the MI attack and corresponds to the state just before the first loss in connectivity when one or several nodes becomes disconnected from the network.  
(a) Cumulative probability density function of normalised channel flow velocity, 
$\tilde{v} = \frac{\left|\bar v_{ij}\right|}{v_0}$ where $v_0=\frac{R^2}{8\eta}\nabla P$ is the normalisation velocity scale. 
(b,c) Polar plots of the angular distribution of channel flow velocity. The angular coordinate corresponds to the channel orientation with respect to the flow direction and the radial coordinate represents the magnitude of the average normalised velocity of channels at this orientation.
(d,e) The flow velocity field after the MI and mb attacks, respectively. Channels are coloured as a function of their normalised flow velocity. All but the fastest channel velocity for the MI attack ($\tilde{v} = 5.4$, identified as $\tilde{v}_{max}$ in grey in (d)) are presented in (a) and the colorbar of (d)} 
\label{fig3:attack_vel}    
\end{figure}

As illustrated in figure \ref{fig3:attack_vel} (a)-(c), each of the four attack types alter the distribution of channel flow velocities from that of the undamaged state. 
Most notably, the major attacks introduce the greatest degree of heterogeneity in the flow field as evidenced by a broader distribution of flow rates. These distributions highlight a stark difference between the fastest flow rates carried by the minority of channels and the flow rates of the majority which are significantly slower than those of the undamaged networks. In the particular case of MI attacks, all fluid pathways through the network are constrained to converge toward and through the limited number of channel `highways' (see grey channel in Fig. \ref{fig3:attack_vel}(d) for example) that enable the network to percolate (and also sustain the fastest flow rates). As a result, the fluid is forced to experience a wide range of flow velocities on its path to the outlet. A similar phenomena is observed for MB attacks as we increase $\mathcal{P}$ and approach the percolation limit. 

The minor attacks eliminate the slowest channels which tend to be transverse to the macroscopic flow, narrowing the distribution of channel velocities and reducing the network tortuosity.

Heterogeneous flow fields, a combination of fast and slow transport rates within the system, preferential flows, and convergent and divergent flow paths, are all conditions that are known to alter the nature of the transfer dynamic and contribute to anomalous transport \cite{bijeljic2011signature, berkowitz2006modeling, edery2016structural, lester2013chaotic, lester2016chaotic}. Whether these factors, introduced to differing degrees by the different network attacks can also impact macroscopic transfer properties like transfer efficiency, is yet to be seen.    

\subsection{Transfer dynamics of clogged networks in regime II}

\begin{figure}
\centering
\includegraphics[width=\columnwidth]{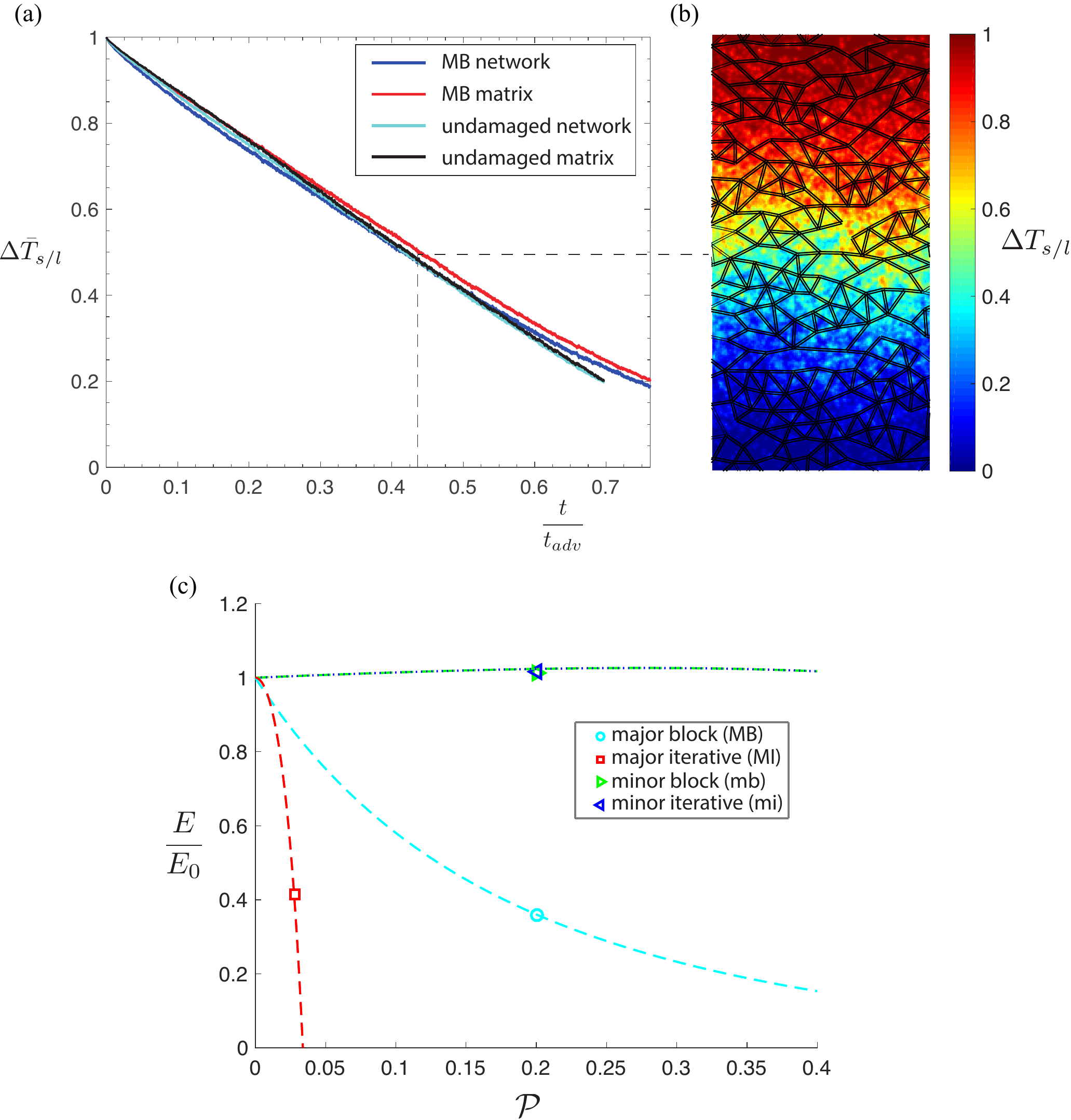}
\caption{Thermal response of an attacked network in regime II. The example results are from a cooling experiment performed using the tracer method after a major block attack ($\mathcal{P}=20\%$, $\phi=0.41$, $T=2.27$, $Pe=18$) (a) Normalised average temperature variation of the matrix (red) and liquid in the channel network (blue) versus time normalised by $t_{adv}$. 
(b) Snapshot of the system's temperature field at $t_{50}$ where $\Delta \bar T_s(t=t_{50})=0.5$
(c) Normalised transfer efficiency results for regime II (markers) versus proportion of clogged channels for each network attack. Dashed lines represent predictions of: Eq. \ref{Eq:E_II_most_attacks} for MB ($b=2.45$), mi and mb attacks ($b=-0.58$); and Eq. \ref{Eq:E_II_MI_attack} for MI attacks ($\mathcal{P}_c=0.0406$)
}
\label{fig4:attack_regII}    
\end{figure}

Figure \ref{fig4:attack_regII} illustrates a typical thermal response for an attacked network subjected to the lowest of the pressure gradients applied. Once again, thermal behaviours congruent with those witnessed in regime II for the undamaged networks are observed.
 In particular, the average temperature of the matrix is similar to that of the liquid in the network throughout the cooling process and both decay linearly with time. Despite the attacks altering the Darcy flow velocity, when each cooling time is normalised by its corresponding advection time scale $t_{adv}$, the transfer dynamics of the attacked networks closely match that of the undamaged network. The temperature fields also demonstrate very little perturbation in the way the thermal fronts progress through the system following the attacks, as the diffusion in the matrix keeps pace with the advection in the channels. Together the results demonstrate signature regime II behaviours and cooling times that are governed by advection, with the efficiency in regime II:
\begin{equation}
E_{II} \approx Pe
\end{equation} 
for all attacks. 
 
Thus in regime II, the vulnerability of a system's transfer efficiency to the different attacks scales with their effect on the permeability. Following from the permeability relationships developed in \cite{rognon2016vulnerability}, this vulnerability, represented by efficiency following an attack relative to that of the undamaged network $E_0$ (at the same $\nabla P$ and  $\eta$) can be expressed as:
  
 \begin{equation}\label{Eq:E_II_most_attacks}
  \left(\frac{E}{E_0}\right)_{II} \approx \frac{1-\mathcal{P}}{\left(1+b \mathcal{P} \right) ^{2}}
 \end{equation}
for all attack types except the major iterative attacks which follow:

  \begin{equation}\label{Eq:E_II_MI_attack}
 \left(\frac{E_{MI}}{E_0}\right)_{II}\approx 2 -e^{\left( \frac{\mathcal{P}}{\mathcal{P}_c}\right)^2}
 \end{equation}
$b$ is a numerical constant that can be positive or negative and  $\mathcal{P}_c$ can be related to the proportion of clogged channels leading to a complete loss of permeability. 
Fig. \ref{fig4:attack_regII}(c) illustrates that the transfer efficiency is most vulnerable to major attacks, with both types leading to significant losses. Most notably, major iterative attacks lead to a dramatic decrease in efficiency even when a relatively small proportion of channels are clogged. On the other hand, the slight improvement in permeability observed following both types of minor attacks translates to a proportionate increase in transfer efficiency. Unlike major attacks, the efficiency is relatively insensitive to the extent of both minor attacks. This stems from differences in the way the network microstructure and flow field evolves as it is attacked with increasing $\mathcal{P}$. In particular, minor attacks reduce tortuosity at a rate that almost balances the effect of the reduction in porosity on permeability with $\mathcal{P}$.

\subsection{Transfer dynamics of clogged networks in regime III}
\begin{figure}
\centering
\includegraphics[width=\columnwidth]{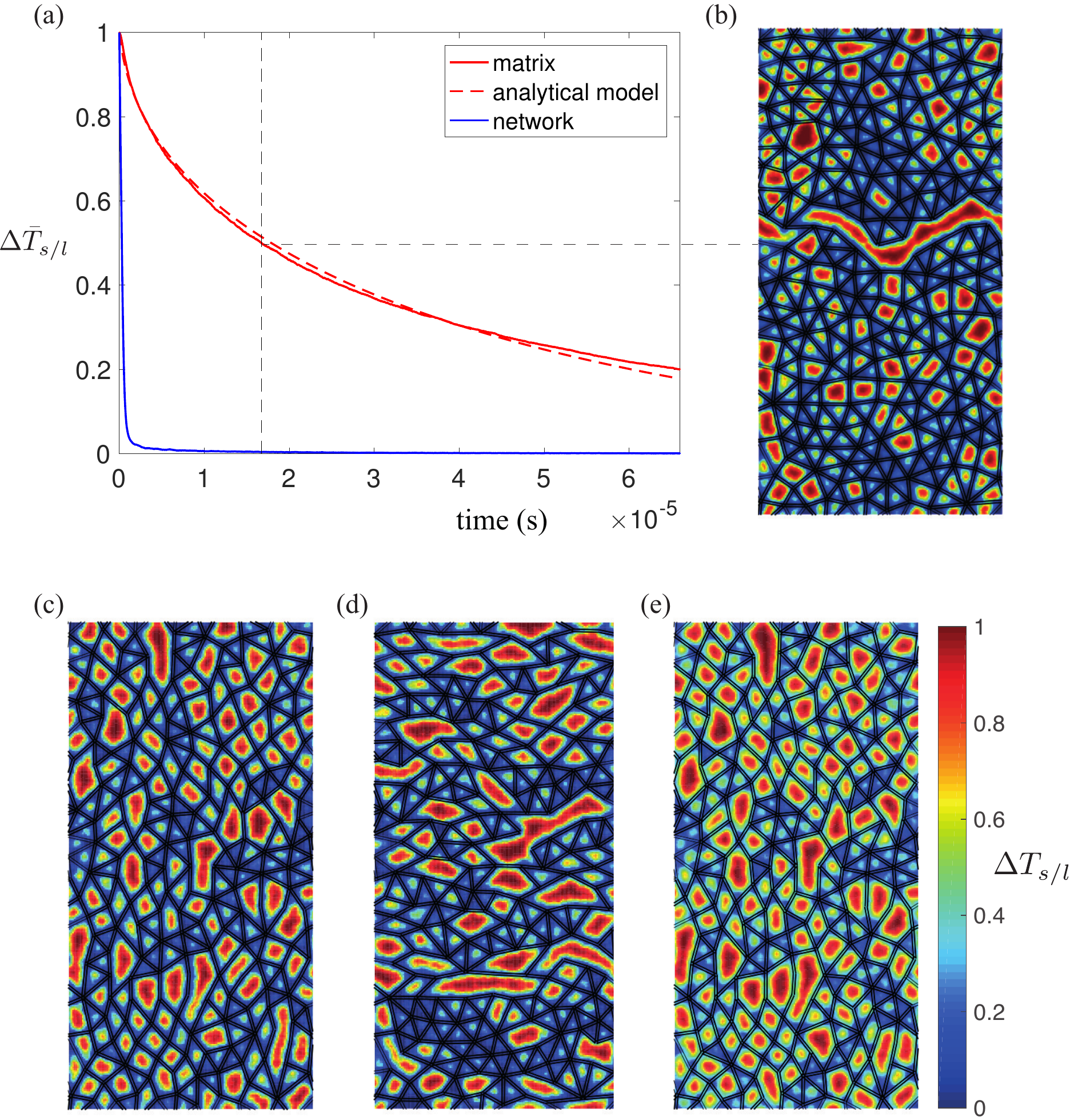}
\caption{Thermal responses of attacked networks in regime III following the same attacks illustrated in Fig. \ref{fig3:attack_vel} (a) Example normalised average temperature variation of the matrix (red) and liquid in the channel network (blue) versus dimensionless time for a MI attack ($\mathcal{P}=20\%$, $\phi=0.48$, $T$, $Pe=8.84\times 10^5$). The cooling dynamics for the matrix obtained using the tracer method (red) are compared with the prediction (red dashed) of the analytical model (Eq. \ref{eq:dirichlet_2d}). Best fit was achieved with  $\ell_e=0.58\bar{l}$, truncating the sum to $n, m <= 19$.
Snapshots of the temperature field at $t_{50}$ (where $\Delta \bar T_s(t=t_{50})=0.5$) for systems subjected to:  (b) the same MI attack depicted in (a); and for a (c) mi; (d) MB; and (e) mb attack.}
\label{fig5:attack_regIII}    
\end{figure}

All attacked systems subjected to the highest of the pressure gradients tested demonstrate thermal responses consistent with regime III. Figure \ref{fig5:attack_regIII}(a), illustrates the typical response whereby the temporal evolution of average temperatures in the matrix and liquid differ significantly. The temperature fields for each attack reflect this difference and as with the undamaged systems, matrix islands quickly emerge, albeit of different size, shape and spatial distribution (see examples in Fig. \ref{fig5:attack_regIII}(b-e)). 

\begin{figure}
\centering
\includegraphics[width=0.75\columnwidth]{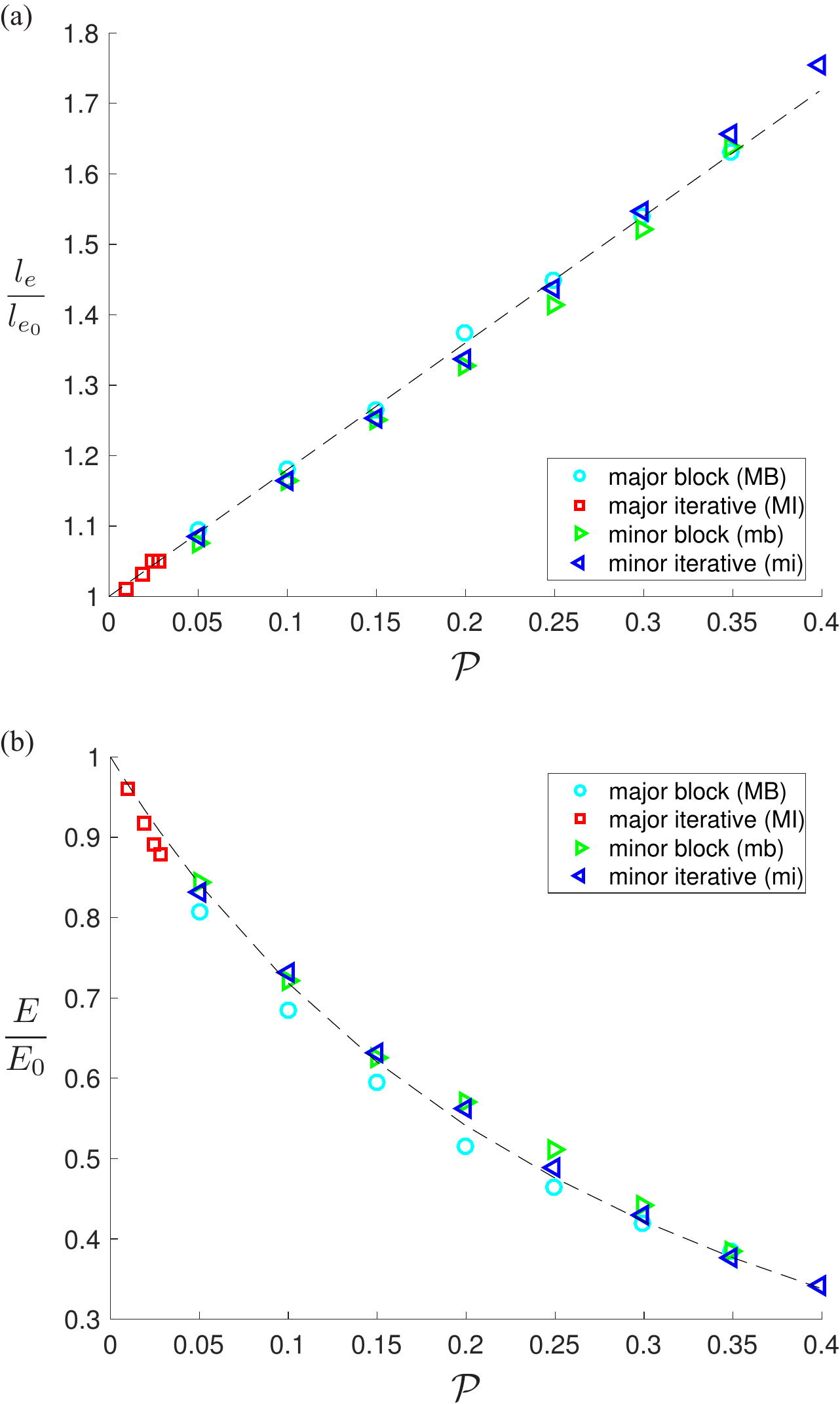}
\caption{ Effects of attacking the network on the transfer in regime III. (a) Markers represent normalised effective length of matrix islands (determined from the thermal responses in regime III) versus proportion of attacked channels for each attack type. Dashed line represents line of best fit $\frac{l_e}{l_{e0}} = 1.8\mathcal{P} + 1$ (b) Normalised transfer efficiency results for regime III versus proportion of clogged channels for each network attack (markers) compared to model prediction (Eq. \ref{Eq:Enorm_III}) (dashed line).}
\label{fig6:attack_l_E_regIII}    
\end{figure}

The presence of matrix islands suggests the matrix cooling dynamic is again controlled by the process of `micro-diffusion', with each attack simply changing the $l_e$ of the islands. To test this hypothesis, the analytical model (\ref{eq:dirichlet_2d}) was fit to tracer method results for each attacked system by altering $l_e$. In every case, the analytical model closely reproduced the cooling dynamics (with $R^2 \geq 0.98$), confirming the hypothesis. An example fit is illustrated in Fig. \ref{fig5:attack_regIII}(a).
The effective island lengths yielded by this processes for every attack are normalised by the effective island length of the undamaged network, $l_0$ and plotted as a function of the proportion of attacked channels, $\mathcal{P}$ in Fig. \ref{fig6:attack_l_E_regIII}. The normalised effective lengths increase linearly with $\mathcal{P}$ and can be captured by the relationship: 
\begin{equation}\label{Eq:l_eff_P}
l_e = l_0 \left(1.8\mathcal{P}+1\right)
\end{equation}
This relationship reflects the generic effect of all attacks considered on the network geometry, as removing a channel between neighbouring matrix islands roughly doubles the size of the island.      

The analytical model (Eq. \ref{eq:dirichlet_2d}) highlights that in regime III, the transfer from the matrix to the fractures involves a spectrum of rates characterised by a series of time scales $t_{n,m}$. 
This spectrum reflects a basic physical process: initially cooling at the island boundary is very quick as the distance heat needs to diffuse to reach the channels is very short; after some time, the outer ring of the island has cooled down and heat must travel a larger distance from the island core to the boundary, taking more time.
$t_{1,1}=\frac{\ell_e^2}{D_o}\frac{1}{2\pi^2}$ corresponds to the slowest rate and governs the cooling dynamic at late times $t$. 

The rates governing the cooling time, $t_{80}$ in regime III can be approximated by:
$t_{80}=\frac{1}{1.46 \pi^2}\frac{\ell_e^2}{D_0}$
\noindent This leads to an efficiency for regime III of:
\begin{eqnarray}\label{Eq:E_III}
E_{III} = 1.46\pi ^2 \lambda ^2  
\end{eqnarray}

\noindent which agrees with the efficiency limit $E_{\infty}$ of every system simulated.

Utilising Eq. \ref{Eq:l_eff_P}, the normalised efficiency in regime III can be expressed as:
\begin{eqnarray}\label{Eq:Enorm_III}
\left(\frac{E}{E_0}\right)_{III}= \frac{1}{\left(1.8\mathcal{P}+1\right)^2}
\end{eqnarray}

This equation holds for all attacked systems as well as the undamaged system ($\mathcal{P}=0$, $\ell_e=\ell_0$), noting no dependence on the type of attack, the direction of the clogged channels, or the degree of heterogeneity in the network geometry, topology or flow field.

\subsection{Transition from regime II to regime III}

 As illustrated in Fig. \ref{fig2:undam}(a), the following interpolation function captures the transfer efficiency results for the undamaged systems in all three regimes: 
\bee \label{Eq:model_E}
E=1+\frac{Pe}{\left(1+\frac{Pe}{E_{\infty}}\right)}
\eee

The function reflects a transition from regime II to regime III when $Pe=E_{\infty}=E_{III}$.

In light of Eq. \ref{Eq:E_III}, it is inferred that the increase in the effective matrix island size ($l_e$) caused by network attacks would reduce the maximum transfer efficiency limit and shift this transition to the left (to coincide with lower $Pe$ values). At the same time, we have observed that the different types of attacks can yield very different effects on the flow ($q$) and thus for a given $\nabla P$, $Pe$ may decrease or increase following attack. 
Together this implies that an attack may cause an undamaged system to suddenly move from regime II to regime III with immediate consequences for the efficiency (and its responsiveness to attempts to improve it.)

\section{Conclusions}
This study quantifies the strong effect attacks to a hydraulic network have on their ability to aid transfers between the flowing fluids they confine and the surrounding matrix. 

Each of the different types of network attacks considered pose a different threat to a system's transfer efficiency. A unifying framework is presented to rationalise and predict this vulnerability. 

The framework accounts for the impact of the attacks on the underlying physical mechanisms found to govern the transfer dynamics in both undamaged and attacked systems. These mechanisms are used to delineate regimes of efficiency. We focus on regime II and III where advection and micro-diffusion govern, respectively. 

The results suggest that in regime II, the vulnerability of the transfer efficiency is entirely controlled by the effect of the attack on the network permeability. As a result of their disparate influences on the tortuosity and porosity; the permeability, and hence efficiency, is very sensitive to both the type and extent of the attack. 

By contrast, in regime III, the transfer efficiency is relatively insensitive to the type of network attack. Rather, the loss in efficiency can be entirely predicted by accounting for the increase in effective island size which scales with the proportion of clogged channels.

Furthermore, network attacks shift the transition between regime II and III and reduce the maximum efficiency achievable in regime III. 

This framework serves as a useful basis to further analyse the effects of clogging on realistic networks featuring greater heterogeneities, poroelastic channel walls, as well as attacks that partially clog channels thereby constricting their effective apertures.

\appendix
\section{Appendix}

\begin{figure}
\centering
\includegraphics[width=\columnwidth]{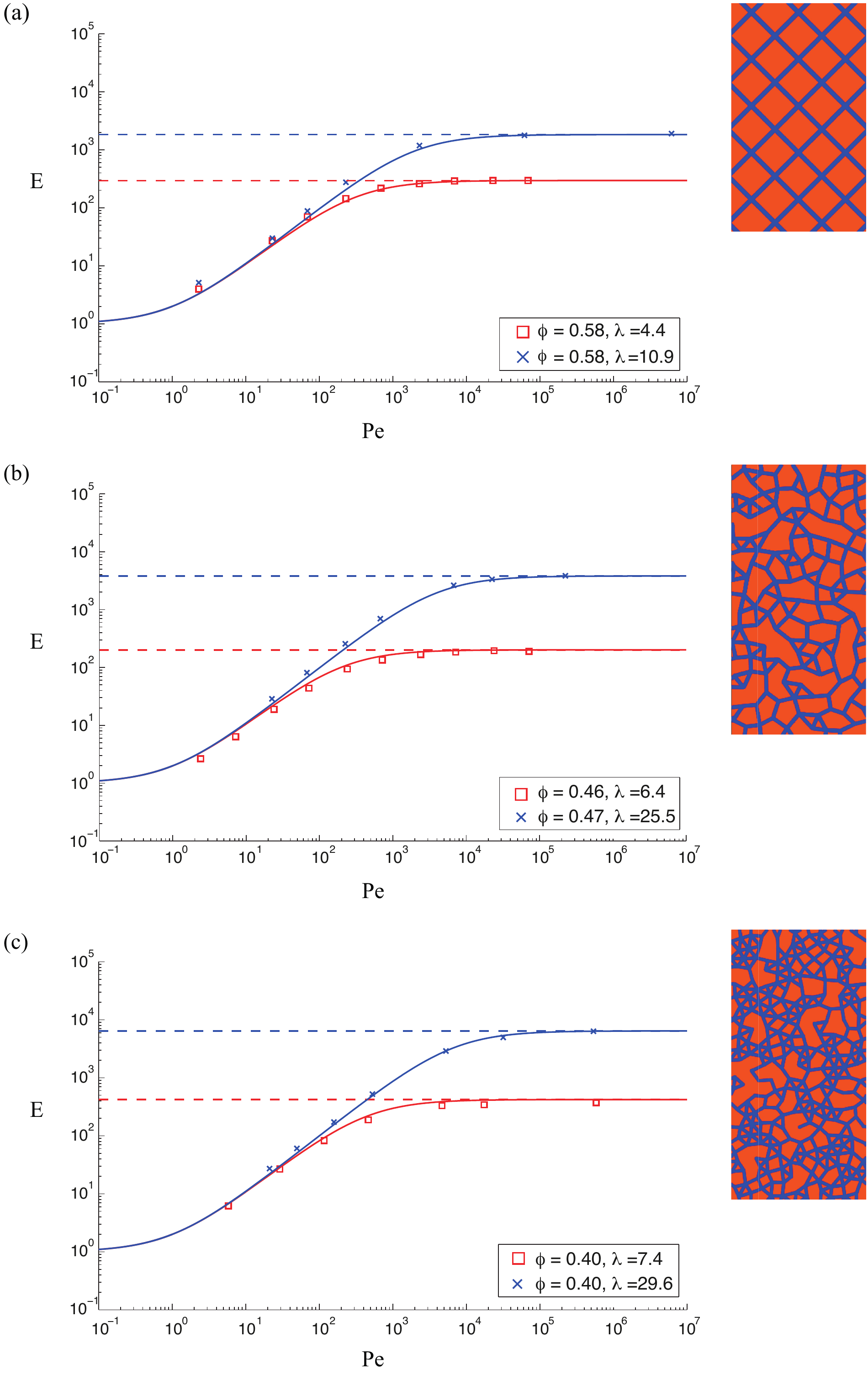}
\caption{Transfer efficiency ($E$) versus P\'eclet number ($Pe$) for systems featuring (a) Lattice (b) K-nearest ($k=3$) and (c) Epsilon ($\epsilon = 11$) hydraulic network topologies. For each type of topology, the following is presented: series of results from tracer method simulations (markers), the transfer efficiency limits obtained from $Pe_{\infty}$ simulations (dashed lines), predictions of the analytical model \ref{Eq:model_E} (solid lines), and an example schematic of each type of topology (right inset).}
\label{fig7:appendix_alltopo}    
\end{figure}

Figure \ref{fig2:undam} and \ref{fig7:appendix_alltopo} illustrates the role of variations in the network topology on the transfer properties. A series of cooling tests were performed using the tracer method (described in text) on a number of different systems where network topology, geometry ($R$), applied pressure gradient ($\nabla P$) and diffusivity ($D_0$) are varied. Each system is characterised by one of the four types of network topologies it features: Gabriel, Lattice, K-nearest or Epsilon network; plus a unique combination of P\'eclet number $Pe$, porosity $\phi$, and dimensionless island size $\lambda$.  
The results are grouped by topology in Fig. \ref{fig7:appendix_alltopo} with Gabriel network results presented in text in Fig. \ref{fig2:undam}. The results highlight that regardless of the network topology, the transfer properties across all regimes can be predicted by the analytical model \ref{Eq:model_E}. The effect of the topological variations are captured by $Pe$ and $\lambda$ via their impact on the flow rate, $q$ and effective island size, $\l_e$.  

%



\begin{thebibliography}{47}%
\makeatletter
\providecommand \@ifxundefined [1]{%
 \@ifx{#1\undefined}
}%
\providecommand \@ifnum [1]{%
 \ifnum #1\expandafter \@firstoftwo
 \else \expandafter \@secondoftwo
 \fi
}%
\providecommand \@ifx [1]{%
 \ifx #1\expandafter \@firstoftwo
 \else \expandafter \@secondoftwo
 \fi
}%
\providecommand \natexlab [1]{#1}%
\providecommand \enquote  [1]{``#1''}%
\providecommand \bibnamefont  [1]{#1}%
\providecommand \bibfnamefont [1]{#1}%
\providecommand \citenamefont [1]{#1}%
\providecommand \href@noop [0]{\@secondoftwo}%
\providecommand \href [0]{\begingroup \@sanitize@url \@href}%
\providecommand \@href[1]{\@@startlink{#1}\@@href}%
\providecommand \@@href[1]{\endgroup#1\@@endlink}%
\providecommand \@sanitize@url [0]{\catcode `\\12\catcode `\$12\catcode
  `\&12\catcode `\#12\catcode `\^12\catcode `\_12\catcode `\%12\relax}%
\providecommand \@@startlink[1]{}%
\providecommand \@@endlink[0]{}%
\providecommand \url  [0]{\begingroup\@sanitize@url \@url }%
\providecommand \@url [1]{\endgroup\@href {#1}{\urlprefix }}%
\providecommand \urlprefix  [0]{URL }%
\providecommand \Eprint [0]{\href }%
\providecommand \doibase [0]{http://dx.doi.org/}%
\providecommand \selectlanguage [0]{\@gobble}%
\providecommand \bibinfo  [0]{\@secondoftwo}%
\providecommand \bibfield  [0]{\@secondoftwo}%
\providecommand \translation [1]{[#1]}%
\providecommand \BibitemOpen [0]{}%
\providecommand \bibitemStop [0]{}%
\providecommand \bibitemNoStop [0]{.\EOS\space}%
\providecommand \EOS [0]{\spacefactor3000\relax}%
\providecommand \BibitemShut  [1]{\csname bibitem#1\endcsname}%
\let\auto@bib@innerbib\@empty
\bibitem [{\citenamefont {Berkowitz}(2002)}]{berkowitz2002characterizing}%
  \BibitemOpen
  \bibfield  {author} {\bibinfo {author} {\bibfnamefont {B.}~\bibnamefont
  {Berkowitz}},\ }\href@noop {} {\bibfield  {journal} {\bibinfo  {journal}
  {Advances in water resources}\ }\textbf {\bibinfo {volume} {25}},\ \bibinfo
  {pages} {861} (\bibinfo {year} {2002})}\BibitemShut {NoStop}%
\bibitem [{\citenamefont {Geiger}\ and\ \citenamefont
  {Emmanuel}(2010)}]{geiger2010non}%
  \BibitemOpen
  \bibfield  {author} {\bibinfo {author} {\bibfnamefont {S.}~\bibnamefont
  {Geiger}}\ and\ \bibinfo {author} {\bibfnamefont {S.}~\bibnamefont
  {Emmanuel}},\ }\href@noop {} {\bibfield  {journal} {\bibinfo  {journal}
  {Water Resources Research}\ }\textbf {\bibinfo {volume} {46}} (\bibinfo
  {year} {2010})}\BibitemShut {NoStop}%
\bibitem [{\citenamefont {Jensen}\ and\ \citenamefont
  {Zwieniecki}(2013)}]{jensen2013physical}%
  \BibitemOpen
  \bibfield  {author} {\bibinfo {author} {\bibfnamefont {K.~H.}\ \bibnamefont
  {Jensen}}\ and\ \bibinfo {author} {\bibfnamefont {M.~A.}\ \bibnamefont
  {Zwieniecki}},\ }\href@noop {} {\bibfield  {journal} {\bibinfo  {journal}
  {Physical review letters}\ }\textbf {\bibinfo {volume} {110}},\ \bibinfo
  {pages} {018104} (\bibinfo {year} {2013})}\BibitemShut {NoStop}%
\bibitem [{\citenamefont {Corson}(2010)}]{corson2010fluctuations}%
  \BibitemOpen
  \bibfield  {author} {\bibinfo {author} {\bibfnamefont {F.}~\bibnamefont
  {Corson}},\ }\href@noop {} {\bibfield  {journal} {\bibinfo  {journal}
  {Physical Review Letters}\ }\textbf {\bibinfo {volume} {104}},\ \bibinfo
  {pages} {048703} (\bibinfo {year} {2010})}\BibitemShut {NoStop}%
\bibitem [{\citenamefont {Katifori}\ \emph {et~al.}(2010)\citenamefont
  {Katifori}, \citenamefont {Sz{\"o}ll{\H{o}}si},\ and\ \citenamefont
  {Magnasco}}]{katifori2010damage}%
  \BibitemOpen
  \bibfield  {author} {\bibinfo {author} {\bibfnamefont {E.}~\bibnamefont
  {Katifori}}, \bibinfo {author} {\bibfnamefont {G.~J.}\ \bibnamefont
  {Sz{\"o}ll{\H{o}}si}}, \ and\ \bibinfo {author} {\bibfnamefont {M.~O.}\
  \bibnamefont {Magnasco}},\ }\href@noop {} {\bibfield  {journal} {\bibinfo
  {journal} {Physical Review Letters}\ }\textbf {\bibinfo {volume} {104}},\
  \bibinfo {pages} {048704} (\bibinfo {year} {2010})}\BibitemShut {NoStop}%
\bibitem [{\citenamefont {Sack}\ and\ \citenamefont
  {Holbrook}(2006)}]{sack2006leaf}%
  \BibitemOpen
  \bibfield  {author} {\bibinfo {author} {\bibfnamefont {L.}~\bibnamefont
  {Sack}}\ and\ \bibinfo {author} {\bibfnamefont {N.~M.}\ \bibnamefont
  {Holbrook}},\ }\href@noop {} {\bibfield  {journal} {\bibinfo  {journal}
  {Annu. Rev. Plant Biol.}\ }\textbf {\bibinfo {volume} {57}},\ \bibinfo
  {pages} {361} (\bibinfo {year} {2006})}\BibitemShut {NoStop}%
\bibitem [{\citenamefont {Pence}(2010)}]{pence2010simplicity}%
  \BibitemOpen
  \bibfield  {author} {\bibinfo {author} {\bibfnamefont {D.}~\bibnamefont
  {Pence}},\ }\href@noop {} {\bibfield  {journal} {\bibinfo  {journal}
  {Experimental Thermal and Fluid Science}\ }\textbf {\bibinfo {volume} {34}},\
  \bibinfo {pages} {474} (\bibinfo {year} {2010})}\BibitemShut {NoStop}%
\bibitem [{\citenamefont {Miller}\ \emph {et~al.}(2012)\citenamefont {Miller},
  \citenamefont {Stevens}, \citenamefont {Yang}, \citenamefont {Baker},
  \citenamefont {Nguyen}, \citenamefont {Cohen}, \citenamefont {Toro},
  \citenamefont {Chen}, \citenamefont {Galie}, \citenamefont {Yu} \emph
  {et~al.}}]{miller2012rapid}%
  \BibitemOpen
  \bibfield  {author} {\bibinfo {author} {\bibfnamefont {J.~S.}\ \bibnamefont
  {Miller}}, \bibinfo {author} {\bibfnamefont {K.~R.}\ \bibnamefont {Stevens}},
  \bibinfo {author} {\bibfnamefont {M.~T.}\ \bibnamefont {Yang}}, \bibinfo
  {author} {\bibfnamefont {B.~M.}\ \bibnamefont {Baker}}, \bibinfo {author}
  {\bibfnamefont {D.-H.~T.}\ \bibnamefont {Nguyen}}, \bibinfo {author}
  {\bibfnamefont {D.~M.}\ \bibnamefont {Cohen}}, \bibinfo {author}
  {\bibfnamefont {E.}~\bibnamefont {Toro}}, \bibinfo {author} {\bibfnamefont
  {A.~A.}\ \bibnamefont {Chen}}, \bibinfo {author} {\bibfnamefont {P.~A.}\
  \bibnamefont {Galie}}, \bibinfo {author} {\bibfnamefont {X.}~\bibnamefont
  {Yu}},  \emph {et~al.},\ }\href@noop {} {\bibfield  {journal} {\bibinfo
  {journal} {Nature materials}\ }\textbf {\bibinfo {volume} {11}},\ \bibinfo
  {pages} {768} (\bibinfo {year} {2012})}\BibitemShut {NoStop}%
\bibitem [{\citenamefont {Hu}\ \emph {et~al.}(2014)\citenamefont {Hu},
  \citenamefont {Zhou}, \citenamefont {Zhu}, \citenamefont {Fan},\ and\
  \citenamefont {Zhang}}]{hu2014efficient}%
  \BibitemOpen
  \bibfield  {author} {\bibinfo {author} {\bibfnamefont {L.}~\bibnamefont
  {Hu}}, \bibinfo {author} {\bibfnamefont {H.}~\bibnamefont {Zhou}}, \bibinfo
  {author} {\bibfnamefont {H.}~\bibnamefont {Zhu}}, \bibinfo {author}
  {\bibfnamefont {T.}~\bibnamefont {Fan}}, \ and\ \bibinfo {author}
  {\bibfnamefont {D.}~\bibnamefont {Zhang}},\ }\href@noop {} {\bibfield
  {journal} {\bibinfo  {journal} {Soft matter}\ }\textbf {\bibinfo {volume}
  {10}},\ \bibinfo {pages} {8442} (\bibinfo {year} {2014})}\BibitemShut
  {NoStop}%
\bibitem [{\citenamefont {Guariguata}\ \emph {et~al.}(2012)\citenamefont
  {Guariguata}, \citenamefont {Pascall}, \citenamefont {Gilmer}, \citenamefont
  {Sum}, \citenamefont {Sloan}, \citenamefont {Koh},\ and\ \citenamefont
  {Wu}}]{guariguata2012jamming}%
  \BibitemOpen
  \bibfield  {author} {\bibinfo {author} {\bibfnamefont {A.}~\bibnamefont
  {Guariguata}}, \bibinfo {author} {\bibfnamefont {M.~A.}\ \bibnamefont
  {Pascall}}, \bibinfo {author} {\bibfnamefont {M.~W.}\ \bibnamefont {Gilmer}},
  \bibinfo {author} {\bibfnamefont {A.~K.}\ \bibnamefont {Sum}}, \bibinfo
  {author} {\bibfnamefont {E.~D.}\ \bibnamefont {Sloan}}, \bibinfo {author}
  {\bibfnamefont {C.~A.}\ \bibnamefont {Koh}}, \ and\ \bibinfo {author}
  {\bibfnamefont {D.~T.}\ \bibnamefont {Wu}},\ }\href@noop {} {\bibfield
  {journal} {\bibinfo  {journal} {Physical Review E}\ }\textbf {\bibinfo
  {volume} {86}},\ \bibinfo {pages} {061311} (\bibinfo {year}
  {2012})}\BibitemShut {NoStop}%
\bibitem [{\citenamefont {Campbell}\ and\ \citenamefont
  {Haw}(2010)}]{campbell2010jamming}%
  \BibitemOpen
  \bibfield  {author} {\bibinfo {author} {\bibfnamefont {A.~I.}\ \bibnamefont
  {Campbell}}\ and\ \bibinfo {author} {\bibfnamefont {M.~D.}\ \bibnamefont
  {Haw}},\ }\href@noop {} {\bibfield  {journal} {\bibinfo  {journal} {Soft
  Matter}\ }\textbf {\bibinfo {volume} {6}},\ \bibinfo {pages} {4688} (\bibinfo
  {year} {2010})}\BibitemShut {NoStop}%
\bibitem [{\citenamefont {Freund}(2013)}]{freund2013flow}%
  \BibitemOpen
  \bibfield  {author} {\bibinfo {author} {\bibfnamefont {J.~B.}\ \bibnamefont
  {Freund}},\ }\href@noop {} {\bibfield  {journal} {\bibinfo  {journal}
  {Physics of Fluids}\ }\textbf {\bibinfo {volume} {25}},\ \bibinfo {pages}
  {110807} (\bibinfo {year} {2013})}\BibitemShut {NoStop}%
\bibitem [{\citenamefont {Valdes}\ \emph {et~al.}(2006)\citenamefont {Valdes},
  \citenamefont {Santamarina} \emph {et~al.}}]{valdes2006particle}%
  \BibitemOpen
  \bibfield  {author} {\bibinfo {author} {\bibfnamefont {J.~R.}\ \bibnamefont
  {Valdes}}, \bibinfo {author} {\bibfnamefont {J.~C.}\ \bibnamefont
  {Santamarina}},  \emph {et~al.},\ }\href@noop {} {\bibfield  {journal}
  {\bibinfo  {journal} {SPE Journal}\ }\textbf {\bibinfo {volume} {11}},\
  \bibinfo {pages} {193} (\bibinfo {year} {2006})}\BibitemShut {NoStop}%
\bibitem [{\citenamefont {Valdes}\ and\ \citenamefont
  {Carlos~Santamarina}(2007)}]{valdes2007particle}%
  \BibitemOpen
  \bibfield  {author} {\bibinfo {author} {\bibfnamefont {J.~R.}\ \bibnamefont
  {Valdes}}\ and\ \bibinfo {author} {\bibfnamefont {J.}~\bibnamefont
  {Carlos~Santamarina}},\ }\href@noop {} {\bibfield  {journal} {\bibinfo
  {journal} {Applied physics letters}\ }\textbf {\bibinfo {volume} {90}},\
  \bibinfo {pages} {244101} (\bibinfo {year} {2007})}\BibitemShut {NoStop}%
\bibitem [{\citenamefont {Ivanov}\ and\ \citenamefont
  {Chu}(2008)}]{ivanov2008applications}%
  \BibitemOpen
  \bibfield  {author} {\bibinfo {author} {\bibfnamefont {V.}~\bibnamefont
  {Ivanov}}\ and\ \bibinfo {author} {\bibfnamefont {J.}~\bibnamefont {Chu}},\
  }\href@noop {} {\bibfield  {journal} {\bibinfo  {journal} {Reviews in
  Environmental Science and Bio/Technology}\ }\textbf {\bibinfo {volume} {7}},\
  \bibinfo {pages} {139} (\bibinfo {year} {2008})}\BibitemShut {NoStop}%
\bibitem [{\citenamefont {Baveye}\ \emph {et~al.}(1998)\citenamefont {Baveye},
  \citenamefont {Vandevivere}, \citenamefont {Hoyle}, \citenamefont {DeLeo},\
  and\ \citenamefont {de~Lozada}}]{baveye1998environmental}%
  \BibitemOpen
  \bibfield  {author} {\bibinfo {author} {\bibfnamefont {P.}~\bibnamefont
  {Baveye}}, \bibinfo {author} {\bibfnamefont {P.}~\bibnamefont {Vandevivere}},
  \bibinfo {author} {\bibfnamefont {B.~L.}\ \bibnamefont {Hoyle}}, \bibinfo
  {author} {\bibfnamefont {P.~C.}\ \bibnamefont {DeLeo}}, \ and\ \bibinfo
  {author} {\bibfnamefont {D.~S.}\ \bibnamefont {de~Lozada}},\ }\href@noop {}
  {\bibfield  {journal} {\bibinfo  {journal} {Critical reviews in environmental
  science and technology}\ }\textbf {\bibinfo {volume} {28}},\ \bibinfo {pages}
  {123} (\bibinfo {year} {1998})}\BibitemShut {NoStop}%
\bibitem [{\citenamefont {Rinck-Pfeiffer}\ \emph {et~al.}(2000)\citenamefont
  {Rinck-Pfeiffer}, \citenamefont {Ragusa}, \citenamefont {Sztajnbok},\ and\
  \citenamefont {Vandevelde}}]{rinck2000interrelationships}%
  \BibitemOpen
  \bibfield  {author} {\bibinfo {author} {\bibfnamefont {S.}~\bibnamefont
  {Rinck-Pfeiffer}}, \bibinfo {author} {\bibfnamefont {S.}~\bibnamefont
  {Ragusa}}, \bibinfo {author} {\bibfnamefont {P.}~\bibnamefont {Sztajnbok}}, \
  and\ \bibinfo {author} {\bibfnamefont {T.}~\bibnamefont {Vandevelde}},\
  }\href@noop {} {\bibfield  {journal} {\bibinfo  {journal} {Water Research}\
  }\textbf {\bibinfo {volume} {34}},\ \bibinfo {pages} {2110} (\bibinfo {year}
  {2000})}\BibitemShut {NoStop}%
\bibitem [{\citenamefont {Depreitere}\ \emph {et~al.}(2006)\citenamefont
  {Depreitere}, \citenamefont {Van~Lierde}, \citenamefont {Sloten},
  \citenamefont {Van~Audekercke}, \citenamefont {Van Der~Perre}, \citenamefont
  {Plets},\ and\ \citenamefont {Goffin}}]{depreitere2006mechanics}%
  \BibitemOpen
  \bibfield  {author} {\bibinfo {author} {\bibfnamefont {B.}~\bibnamefont
  {Depreitere}}, \bibinfo {author} {\bibfnamefont {C.}~\bibnamefont
  {Van~Lierde}}, \bibinfo {author} {\bibfnamefont {J.~V.}\ \bibnamefont
  {Sloten}}, \bibinfo {author} {\bibfnamefont {R.}~\bibnamefont
  {Van~Audekercke}}, \bibinfo {author} {\bibfnamefont {G.}~\bibnamefont {Van
  Der~Perre}}, \bibinfo {author} {\bibfnamefont {C.}~\bibnamefont {Plets}}, \
  and\ \bibinfo {author} {\bibfnamefont {J.}~\bibnamefont {Goffin}},\
  }\href@noop {} {\bibfield  {journal} {\bibinfo  {journal} {Journal of
  neurosurgery}\ }\textbf {\bibinfo {volume} {104}},\ \bibinfo {pages} {950}
  (\bibinfo {year} {2006})}\BibitemShut {NoStop}%
\bibitem [{\citenamefont {Shipley}\ and\ \citenamefont
  {Chapman}(2010)}]{shipley2010multiscale}%
  \BibitemOpen
  \bibfield  {author} {\bibinfo {author} {\bibfnamefont {R.~J.}\ \bibnamefont
  {Shipley}}\ and\ \bibinfo {author} {\bibfnamefont {S.~J.}\ \bibnamefont
  {Chapman}},\ }\href@noop {} {\bibfield  {journal} {\bibinfo  {journal}
  {Bulletin of mathematical biology}\ }\textbf {\bibinfo {volume} {72}},\
  \bibinfo {pages} {1464} (\bibinfo {year} {2010})}\BibitemShut {NoStop}%
\bibitem [{\citenamefont {Bebber}\ \emph {et~al.}(2007)\citenamefont {Bebber},
  \citenamefont {Hynes}, \citenamefont {Darrah}, \citenamefont {Boddy},\ and\
  \citenamefont {Fricker}}]{bebber2007biological}%
  \BibitemOpen
  \bibfield  {author} {\bibinfo {author} {\bibfnamefont {D.~P.}\ \bibnamefont
  {Bebber}}, \bibinfo {author} {\bibfnamefont {J.}~\bibnamefont {Hynes}},
  \bibinfo {author} {\bibfnamefont {P.~R.}\ \bibnamefont {Darrah}}, \bibinfo
  {author} {\bibfnamefont {L.}~\bibnamefont {Boddy}}, \ and\ \bibinfo {author}
  {\bibfnamefont {M.~D.}\ \bibnamefont {Fricker}},\ }\href@noop {} {\bibfield
  {journal} {\bibinfo  {journal} {Proceedings of the Royal Society of London B:
  Biological Sciences}\ }\textbf {\bibinfo {volume} {274}},\ \bibinfo {pages}
  {2307} (\bibinfo {year} {2007})}\BibitemShut {NoStop}%
\bibitem [{\citenamefont {Choat}\ \emph {et~al.}(2012)\citenamefont {Choat},
  \citenamefont {Jansen}, \citenamefont {Brodribb}, \citenamefont {Cochard},
  \citenamefont {Delzon}, \citenamefont {Bhaskar}, \citenamefont {Bucci},
  \citenamefont {Feild}, \citenamefont {Gleason}, \citenamefont {Hacke} \emph
  {et~al.}}]{choat2012global}%
  \BibitemOpen
  \bibfield  {author} {\bibinfo {author} {\bibfnamefont {B.}~\bibnamefont
  {Choat}}, \bibinfo {author} {\bibfnamefont {S.}~\bibnamefont {Jansen}},
  \bibinfo {author} {\bibfnamefont {T.~J.}\ \bibnamefont {Brodribb}}, \bibinfo
  {author} {\bibfnamefont {H.}~\bibnamefont {Cochard}}, \bibinfo {author}
  {\bibfnamefont {S.}~\bibnamefont {Delzon}}, \bibinfo {author} {\bibfnamefont
  {R.}~\bibnamefont {Bhaskar}}, \bibinfo {author} {\bibfnamefont {S.~J.}\
  \bibnamefont {Bucci}}, \bibinfo {author} {\bibfnamefont {T.~S.}\ \bibnamefont
  {Feild}}, \bibinfo {author} {\bibfnamefont {S.~M.}\ \bibnamefont {Gleason}},
  \bibinfo {author} {\bibfnamefont {U.~G.}\ \bibnamefont {Hacke}},  \emph
  {et~al.},\ }\href@noop {} {\bibfield  {journal} {\bibinfo  {journal}
  {Nature}\ }\textbf {\bibinfo {volume} {491}},\ \bibinfo {pages} {752}
  (\bibinfo {year} {2012})}\BibitemShut {NoStop}%
\bibitem [{\citenamefont {Loepfe}\ \emph {et~al.}(2007)\citenamefont {Loepfe},
  \citenamefont {Martinez-Vilalta}, \citenamefont {Pi{\~n}ol},\ and\
  \citenamefont {Mencuccini}}]{loepfe2007relevance}%
  \BibitemOpen
  \bibfield  {author} {\bibinfo {author} {\bibfnamefont {L.}~\bibnamefont
  {Loepfe}}, \bibinfo {author} {\bibfnamefont {J.}~\bibnamefont
  {Martinez-Vilalta}}, \bibinfo {author} {\bibfnamefont {J.}~\bibnamefont
  {Pi{\~n}ol}}, \ and\ \bibinfo {author} {\bibfnamefont {M.}~\bibnamefont
  {Mencuccini}},\ }\href@noop {} {\bibfield  {journal} {\bibinfo  {journal}
  {Journal of Theoretical Biology}\ }\textbf {\bibinfo {volume} {247}},\
  \bibinfo {pages} {788} (\bibinfo {year} {2007})}\BibitemShut {NoStop}%
\bibitem [{\citenamefont {Singurindy}\ and\ \citenamefont
  {Berkowitz}(2005)}]{singurindy2005role}%
  \BibitemOpen
  \bibfield  {author} {\bibinfo {author} {\bibfnamefont {O.}~\bibnamefont
  {Singurindy}}\ and\ \bibinfo {author} {\bibfnamefont {B.}~\bibnamefont
  {Berkowitz}},\ }\href@noop {} {\bibfield  {journal} {\bibinfo  {journal}
  {Advances in water resources}\ }\textbf {\bibinfo {volume} {28}},\ \bibinfo
  {pages} {507} (\bibinfo {year} {2005})}\BibitemShut {NoStop}%
\bibitem [{\citenamefont {Andr{\'e}}\ \emph {et~al.}(2006)\citenamefont
  {Andr{\'e}}, \citenamefont {Rabemanana},\ and\ \citenamefont
  {Vuataz}}]{andre2006influence}%
  \BibitemOpen
  \bibfield  {author} {\bibinfo {author} {\bibfnamefont {L.}~\bibnamefont
  {Andr{\'e}}}, \bibinfo {author} {\bibfnamefont {V.}~\bibnamefont
  {Rabemanana}}, \ and\ \bibinfo {author} {\bibfnamefont {F.-D.}\ \bibnamefont
  {Vuataz}},\ }\href@noop {} {\bibfield  {journal} {\bibinfo  {journal}
  {Geothermics}\ }\textbf {\bibinfo {volume} {35}},\ \bibinfo {pages} {507}
  (\bibinfo {year} {2006})}\BibitemShut {NoStop}%
\bibitem [{\citenamefont {B{\"a}chler}\ and\ \citenamefont
  {Kohl}(2005)}]{bachler2005coupled}%
  \BibitemOpen
  \bibfield  {author} {\bibinfo {author} {\bibfnamefont {D.}~\bibnamefont
  {B{\"a}chler}}\ and\ \bibinfo {author} {\bibfnamefont {T.}~\bibnamefont
  {Kohl}},\ }\href@noop {} {\bibfield  {journal} {\bibinfo  {journal}
  {Geophysical Journal International}\ }\textbf {\bibinfo {volume} {161}},\
  \bibinfo {pages} {533} (\bibinfo {year} {2005})}\BibitemShut {NoStop}%
\bibitem [{\citenamefont {Fuerstman}\ \emph {et~al.}(2003)\citenamefont
  {Fuerstman}, \citenamefont {Deschatelets}, \citenamefont {Kane},
  \citenamefont {Schwartz}, \citenamefont {Kenis}, \citenamefont {Deutch},\
  and\ \citenamefont {Whitesides}}]{fuerstman2003solving}%
  \BibitemOpen
  \bibfield  {author} {\bibinfo {author} {\bibfnamefont {M.~J.}\ \bibnamefont
  {Fuerstman}}, \bibinfo {author} {\bibfnamefont {P.}~\bibnamefont
  {Deschatelets}}, \bibinfo {author} {\bibfnamefont {R.}~\bibnamefont {Kane}},
  \bibinfo {author} {\bibfnamefont {A.}~\bibnamefont {Schwartz}}, \bibinfo
  {author} {\bibfnamefont {P.~J.}\ \bibnamefont {Kenis}}, \bibinfo {author}
  {\bibfnamefont {J.~M.}\ \bibnamefont {Deutch}}, \ and\ \bibinfo {author}
  {\bibfnamefont {G.~M.}\ \bibnamefont {Whitesides}},\ }\href@noop {}
  {\bibfield  {journal} {\bibinfo  {journal} {Langmuir}\ }\textbf {\bibinfo
  {volume} {19}},\ \bibinfo {pages} {4714} (\bibinfo {year}
  {2003})}\BibitemShut {NoStop}%
\bibitem [{\citenamefont {Costa}(2006)}]{costa2006permeability}%
  \BibitemOpen
  \bibfield  {author} {\bibinfo {author} {\bibfnamefont {A.}~\bibnamefont
  {Costa}},\ }\href@noop {} {\bibfield  {journal} {\bibinfo  {journal}
  {Geophysical research letters}\ }\textbf {\bibinfo {volume} {33}} (\bibinfo
  {year} {2006})}\BibitemShut {NoStop}%
\bibitem [{\citenamefont {Jivkov}\ \emph {et~al.}(2013)\citenamefont {Jivkov},
  \citenamefont {Hollis}, \citenamefont {Etiese}, \citenamefont {McDonald},\
  and\ \citenamefont {Withers}}]{jivkov2013novel}%
  \BibitemOpen
  \bibfield  {author} {\bibinfo {author} {\bibfnamefont {A.~P.}\ \bibnamefont
  {Jivkov}}, \bibinfo {author} {\bibfnamefont {C.}~\bibnamefont {Hollis}},
  \bibinfo {author} {\bibfnamefont {F.}~\bibnamefont {Etiese}}, \bibinfo
  {author} {\bibfnamefont {S.~A.}\ \bibnamefont {McDonald}}, \ and\ \bibinfo
  {author} {\bibfnamefont {P.~J.}\ \bibnamefont {Withers}},\ }\href@noop {}
  {\bibfield  {journal} {\bibinfo  {journal} {Journal of Hydrology}\ }\textbf
  {\bibinfo {volume} {486}},\ \bibinfo {pages} {246} (\bibinfo {year}
  {2013})}\BibitemShut {NoStop}%
\bibitem [{\citenamefont {Duda}\ \emph {et~al.}(2011)\citenamefont {Duda},
  \citenamefont {Koza},\ and\ \citenamefont {Matyka}}]{duda2011hydraulic}%
  \BibitemOpen
  \bibfield  {author} {\bibinfo {author} {\bibfnamefont {A.}~\bibnamefont
  {Duda}}, \bibinfo {author} {\bibfnamefont {Z.}~\bibnamefont {Koza}}, \ and\
  \bibinfo {author} {\bibfnamefont {M.}~\bibnamefont {Matyka}},\ }\href@noop {}
  {\bibfield  {journal} {\bibinfo  {journal} {Physical Review E}\ }\textbf
  {\bibinfo {volume} {84}},\ \bibinfo {pages} {036319} (\bibinfo {year}
  {2011})}\BibitemShut {NoStop}%
\bibitem [{\citenamefont {Koponen}\ \emph {et~al.}(1996)\citenamefont
  {Koponen}, \citenamefont {Kataja},\ and\ \citenamefont
  {Timonen}}]{koponen1996tortuous}%
  \BibitemOpen
  \bibfield  {author} {\bibinfo {author} {\bibfnamefont {A.}~\bibnamefont
  {Koponen}}, \bibinfo {author} {\bibfnamefont {M.}~\bibnamefont {Kataja}}, \
  and\ \bibinfo {author} {\bibfnamefont {J.}~\bibnamefont {Timonen}},\
  }\href@noop {} {\bibfield  {journal} {\bibinfo  {journal} {Physical Review
  E}\ }\textbf {\bibinfo {volume} {54}},\ \bibinfo {pages} {406} (\bibinfo
  {year} {1996})}\BibitemShut {NoStop}%
\bibitem [{\citenamefont {Rognon}\ \emph {et~al.}(2014)\citenamefont {Rognon},
  \citenamefont {Macaulay}, \citenamefont {Griffani},\ and\ \citenamefont
  {Einav}}]{rognon2014explaining}%
  \BibitemOpen
  \bibfield  {author} {\bibinfo {author} {\bibfnamefont {P.}~\bibnamefont
  {Rognon}}, \bibinfo {author} {\bibfnamefont {M.}~\bibnamefont {Macaulay}},
  \bibinfo {author} {\bibfnamefont {D.}~\bibnamefont {Griffani}}, \ and\
  \bibinfo {author} {\bibfnamefont {I.}~\bibnamefont {Einav}},\ }\href@noop {}
  {\bibfield  {journal} {\bibinfo  {journal} {EPL (Europhysics Letters)}\
  }\textbf {\bibinfo {volume} {108}},\ \bibinfo {pages} {34004} (\bibinfo
  {year} {2014})}\BibitemShut {NoStop}%
\bibitem [{\citenamefont {Xu}\ and\ \citenamefont
  {Pruess}(2001)}]{xu2001modeling}%
  \BibitemOpen
  \bibfield  {author} {\bibinfo {author} {\bibfnamefont {T.}~\bibnamefont
  {Xu}}\ and\ \bibinfo {author} {\bibfnamefont {K.}~\bibnamefont {Pruess}},\
  }\href@noop {} {\bibfield  {journal} {\bibinfo  {journal} {American Journal
  of Science}\ }\textbf {\bibinfo {volume} {301}},\ \bibinfo {pages} {16}
  (\bibinfo {year} {2001})}\BibitemShut {NoStop}%
\bibitem [{\citenamefont {Pruess}\ and\ \citenamefont
  {Narasimhan}(1982)}]{pruess1982practical}%
  \BibitemOpen
  \bibfield  {author} {\bibinfo {author} {\bibfnamefont {K.}~\bibnamefont
  {Pruess}}\ and\ \bibinfo {author} {\bibfnamefont {T.}~\bibnamefont
  {Narasimhan}},\ }\href@noop {} {\emph {\bibinfo {title} {Practical method for
  modeling fluid and heat flow in fractured porous media}}},\ \bibinfo {type}
  {Tech. Rep.}\ (\bibinfo  {institution} {Lawrence Berkeley Lab., CA (USA)},\
  \bibinfo {year} {1982})\BibitemShut {NoStop}%
\bibitem [{\citenamefont {Haggerty}\ and\ \citenamefont
  {Gorelick}(1995)}]{haggerty1995multiple}%
  \BibitemOpen
  \bibfield  {author} {\bibinfo {author} {\bibfnamefont {R.}~\bibnamefont
  {Haggerty}}\ and\ \bibinfo {author} {\bibfnamefont {S.~M.}\ \bibnamefont
  {Gorelick}},\ }\href@noop {} {\bibfield  {journal} {\bibinfo  {journal}
  {Water Resources Research}\ }\textbf {\bibinfo {volume} {31}},\ \bibinfo
  {pages} {2383} (\bibinfo {year} {1995})}\BibitemShut {NoStop}%
\bibitem [{\citenamefont {Carrera}\ \emph {et~al.}(1998)\citenamefont
  {Carrera}, \citenamefont {S{\'a}nchez-Vila}, \citenamefont {Benet},
  \citenamefont {Medina}, \citenamefont {Galarza},\ and\ \citenamefont
  {Guimer{\`a}}}]{carrera1998matrix}%
  \BibitemOpen
  \bibfield  {author} {\bibinfo {author} {\bibfnamefont {J.}~\bibnamefont
  {Carrera}}, \bibinfo {author} {\bibfnamefont {X.}~\bibnamefont
  {S{\'a}nchez-Vila}}, \bibinfo {author} {\bibfnamefont {I.}~\bibnamefont
  {Benet}}, \bibinfo {author} {\bibfnamefont {A.}~\bibnamefont {Medina}},
  \bibinfo {author} {\bibfnamefont {G.}~\bibnamefont {Galarza}}, \ and\
  \bibinfo {author} {\bibfnamefont {J.}~\bibnamefont {Guimer{\`a}}},\
  }\href@noop {} {\bibfield  {journal} {\bibinfo  {journal} {Hydrogeology
  Journal}\ }\textbf {\bibinfo {volume} {6}},\ \bibinfo {pages} {178} (\bibinfo
  {year} {1998})}\BibitemShut {NoStop}%
\bibitem [{\citenamefont {Rognon}\ \emph {et~al.}(2016)\citenamefont {Rognon},
  \citenamefont {Russo}, \citenamefont {Griffani},\ and\ \citenamefont
  {Einav}}]{rognon2016vulnerability}%
  \BibitemOpen
  \bibfield  {author} {\bibinfo {author} {\bibfnamefont {P.}~\bibnamefont
  {Rognon}}, \bibinfo {author} {\bibfnamefont {D.}~\bibnamefont {Russo}},
  \bibinfo {author} {\bibfnamefont {D.}~\bibnamefont {Griffani}}, \ and\
  \bibinfo {author} {\bibfnamefont {I.}~\bibnamefont {Einav}},\ }\href@noop {}
  {\bibfield  {journal} {\bibinfo  {journal} {EPL (Europhysics Letters)}\
  }\textbf {\bibinfo {volume} {113}},\ \bibinfo {pages} {14001} (\bibinfo
  {year} {2016})}\BibitemShut {NoStop}%
\bibitem [{\citenamefont {Walsh}(1981)}]{walsh1981effect}%
  \BibitemOpen
  \bibfield  {author} {\bibinfo {author} {\bibfnamefont {J.}~\bibnamefont
  {Walsh}},\ }in\ \href@noop {} {\emph {\bibinfo {booktitle} {International
  Journal of Rock Mechanics and Mining Sciences \& Geomechanics Abstracts}}},\
  Vol.~\bibinfo {volume} {18}\ (\bibinfo {organization} {Elsevier},\ \bibinfo
  {year} {1981})\ pp.\ \bibinfo {pages} {429--435}\BibitemShut {NoStop}%
\bibitem [{\citenamefont {Latham}\ \emph {et~al.}(2013)\citenamefont {Latham},
  \citenamefont {Xiang}, \citenamefont {Belayneh}, \citenamefont {Nick},
  \citenamefont {Tsang},\ and\ \citenamefont {Blunt}}]{latham2013modelling}%
  \BibitemOpen
  \bibfield  {author} {\bibinfo {author} {\bibfnamefont {J.-P.}\ \bibnamefont
  {Latham}}, \bibinfo {author} {\bibfnamefont {J.}~\bibnamefont {Xiang}},
  \bibinfo {author} {\bibfnamefont {M.}~\bibnamefont {Belayneh}}, \bibinfo
  {author} {\bibfnamefont {H.~M.}\ \bibnamefont {Nick}}, \bibinfo {author}
  {\bibfnamefont {C.-F.}\ \bibnamefont {Tsang}}, \ and\ \bibinfo {author}
  {\bibfnamefont {M.~J.}\ \bibnamefont {Blunt}},\ }\href@noop {} {\bibfield
  {journal} {\bibinfo  {journal} {International Journal of Rock Mechanics and
  Mining Sciences}\ }\textbf {\bibinfo {volume} {57}},\ \bibinfo {pages} {100}
  (\bibinfo {year} {2013})}\BibitemShut {NoStop}%
\bibitem [{\citenamefont {Carey}\ \emph {et~al.}(2015)\citenamefont {Carey},
  \citenamefont {Lei}, \citenamefont {Rougier}, \citenamefont {Mori},\ and\
  \citenamefont {Viswanathan}}]{carey2015fracture}%
  \BibitemOpen
  \bibfield  {author} {\bibinfo {author} {\bibfnamefont {J.~W.}\ \bibnamefont
  {Carey}}, \bibinfo {author} {\bibfnamefont {Z.}~\bibnamefont {Lei}}, \bibinfo
  {author} {\bibfnamefont {E.}~\bibnamefont {Rougier}}, \bibinfo {author}
  {\bibfnamefont {H.}~\bibnamefont {Mori}}, \ and\ \bibinfo {author}
  {\bibfnamefont {H.}~\bibnamefont {Viswanathan}},\ }\href@noop {} {\bibfield
  {journal} {\bibinfo  {journal} {Journal of Unconventional Oil and Gas
  Resources}\ }\textbf {\bibinfo {volume} {11}},\ \bibinfo {pages} {27}
  (\bibinfo {year} {2015})}\BibitemShut {NoStop}%
\bibitem [{\citenamefont {Griffani}\ \emph {et~al.}(2013)\citenamefont
  {Griffani}, \citenamefont {Rognon}, \citenamefont {Metzger},\ and\
  \citenamefont {Einav}}]{griffani2013rotational}%
  \BibitemOpen
  \bibfield  {author} {\bibinfo {author} {\bibfnamefont {D.}~\bibnamefont
  {Griffani}}, \bibinfo {author} {\bibfnamefont {P.}~\bibnamefont {Rognon}},
  \bibinfo {author} {\bibfnamefont {B.}~\bibnamefont {Metzger}}, \ and\
  \bibinfo {author} {\bibfnamefont {I.}~\bibnamefont {Einav}},\ }\href@noop {}
  {\bibfield  {journal} {\bibinfo  {journal} {Physics of Fluids}\ }\textbf
  {\bibinfo {volume} {25}},\ \bibinfo {pages} {093301} (\bibinfo {year}
  {2013})}\BibitemShut {NoStop}%
\bibitem [{\citenamefont {Griffani}\ \emph {et~al.}(2014)\citenamefont
  {Griffani}, \citenamefont {Rognon},\ and\ \citenamefont
  {Einav}}]{griffani2014transfer}%
  \BibitemOpen
  \bibfield  {author} {\bibinfo {author} {\bibfnamefont {D.}~\bibnamefont
  {Griffani}}, \bibinfo {author} {\bibfnamefont {P.}~\bibnamefont {Rognon}}, \
  and\ \bibinfo {author} {\bibfnamefont {I.}~\bibnamefont {Einav}},\
  }\href@noop {} {\bibfield  {journal} {\bibinfo  {journal} {EPL (Europhysics
  Letters)}\ }\textbf {\bibinfo {volume} {106}},\ \bibinfo {pages} {64002}
  (\bibinfo {year} {2014})}\BibitemShut {NoStop}%
\bibitem [{\citenamefont {Schirmacher}(2015)}]{schirmacher2015random}%
  \BibitemOpen
  \bibfield  {author} {\bibinfo {author} {\bibfnamefont {W.}~\bibnamefont
  {Schirmacher}},\ }in\ \href@noop {} {\emph {\bibinfo {booktitle} {Theory of
  Liquids and Other Disordered Media}}}\ (\bibinfo  {publisher} {Springer},\
  \bibinfo {year} {2015})\ pp.\ \bibinfo {pages} {45--59}\BibitemShut {NoStop}%
\bibitem [{\citenamefont {Bijeljic}\ \emph {et~al.}(2011)\citenamefont
  {Bijeljic}, \citenamefont {Mostaghimi},\ and\ \citenamefont
  {Blunt}}]{bijeljic2011signature}%
  \BibitemOpen
  \bibfield  {author} {\bibinfo {author} {\bibfnamefont {B.}~\bibnamefont
  {Bijeljic}}, \bibinfo {author} {\bibfnamefont {P.}~\bibnamefont
  {Mostaghimi}}, \ and\ \bibinfo {author} {\bibfnamefont {M.~J.}\ \bibnamefont
  {Blunt}},\ }\href@noop {} {\bibfield  {journal} {\bibinfo  {journal}
  {Physical review letters}\ }\textbf {\bibinfo {volume} {107}},\ \bibinfo
  {pages} {204502} (\bibinfo {year} {2011})}\BibitemShut {NoStop}%
\bibitem [{\citenamefont {Berkowitz}\ \emph {et~al.}(2006)\citenamefont
  {Berkowitz}, \citenamefont {Cortis}, \citenamefont {Dentz},\ and\
  \citenamefont {Scher}}]{berkowitz2006modeling}%
  \BibitemOpen
  \bibfield  {author} {\bibinfo {author} {\bibfnamefont {B.}~\bibnamefont
  {Berkowitz}}, \bibinfo {author} {\bibfnamefont {A.}~\bibnamefont {Cortis}},
  \bibinfo {author} {\bibfnamefont {M.}~\bibnamefont {Dentz}}, \ and\ \bibinfo
  {author} {\bibfnamefont {H.}~\bibnamefont {Scher}},\ }\href@noop {}
  {\bibfield  {journal} {\bibinfo  {journal} {Reviews of Geophysics}\ }\textbf
  {\bibinfo {volume} {44}} (\bibinfo {year} {2006})}\BibitemShut {NoStop}%
\bibitem [{\citenamefont {Edery}\ \emph {et~al.}(2016)\citenamefont {Edery},
  \citenamefont {Geiger},\ and\ \citenamefont
  {Berkowitz}}]{edery2016structural}%
  \BibitemOpen
  \bibfield  {author} {\bibinfo {author} {\bibfnamefont {Y.}~\bibnamefont
  {Edery}}, \bibinfo {author} {\bibfnamefont {S.}~\bibnamefont {Geiger}}, \
  and\ \bibinfo {author} {\bibfnamefont {B.}~\bibnamefont {Berkowitz}},\
  }\href@noop {} {\bibfield  {journal} {\bibinfo  {journal} {Water Resources
  Research}\ }\textbf {\bibinfo {volume} {52}},\ \bibinfo {pages} {5634}
  (\bibinfo {year} {2016})}\BibitemShut {NoStop}%
\bibitem [{\citenamefont {Lester}\ \emph {et~al.}(2013)\citenamefont {Lester},
  \citenamefont {Metcalfe},\ and\ \citenamefont {Trefry}}]{lester2013chaotic}%
  \BibitemOpen
  \bibfield  {author} {\bibinfo {author} {\bibfnamefont {D.}~\bibnamefont
  {Lester}}, \bibinfo {author} {\bibfnamefont {G.}~\bibnamefont {Metcalfe}}, \
  and\ \bibinfo {author} {\bibfnamefont {M.}~\bibnamefont {Trefry}},\
  }\href@noop {} {\bibfield  {journal} {\bibinfo  {journal} {Physical review
  letters}\ }\textbf {\bibinfo {volume} {111}},\ \bibinfo {pages} {174101}
  (\bibinfo {year} {2013})}\BibitemShut {NoStop}%
\bibitem [{\citenamefont {Lester}\ \emph {et~al.}(2016)\citenamefont {Lester},
  \citenamefont {Dentz},\ and\ \citenamefont {Le~Borgne}}]{lester2016chaotic}%
  \BibitemOpen
  \bibfield  {author} {\bibinfo {author} {\bibfnamefont {D.~R.}\ \bibnamefont
  {Lester}}, \bibinfo {author} {\bibfnamefont {M.}~\bibnamefont {Dentz}}, \
  and\ \bibinfo {author} {\bibfnamefont {T.}~\bibnamefont {Le~Borgne}},\
  }\href@noop {} {\bibfield  {journal} {\bibinfo  {journal} {Journal of Fluid
  Mechanics}\ }\textbf {\bibinfo {volume} {803}},\ \bibinfo {pages} {144}
  (\bibinfo {year} {2016})}\BibitemShut {NoStop}%
\end{thebibliography}

%

\end{document}